\documentclass[ showpacs, preprintnumbers, amsmath, amssymb,twocolumn]{revtex4}

\usepackage{graphicx}
\usepackage{dcolumn}
\usepackage{bm}
\usepackage{subfigure}

\begin{document}

\preprint{...}

\title{One dimensional chain of quantum molecule motors as a mathematical physics model for muscle fibre}

\author{Tieyan Si}
\affiliation{Max-Planck-Institute for the Physics of Complex Systems. Nothnitzer Street 38, D-01187 Dresden, Germany}

\date{\today}

\begin{abstract}

A quantum chain model of many molecule motors is proposed as a mathematical physics theory on 
the microscopic modeling of classical force-velocity relation and tension 
transients of muscle fibre. We proposed quantum many-particle Hamiltonian to predict the force-velocity relation for the slow release of muscle fibre 
which has no empirical relation yet, it is much more complicate than hyperbolic relation. Using the same Hamiltonian, we predicted the mathematical force-velocity relation when the muscle is stimulated by alternative electric current. The discrepancy between input electric frequency and the muscle oscillation frequency has a physical understanding by Doppler effect in this quantum chain model. Further more, we apply quantum physics phenomena to explore the tension time course of cardiac muscle and insect flight muscle. 
Most of the experimental tension transients curves found their correspondence in the theoretical output of quantum two-level and three-level model. 
Mathematically modeling electric stimulus as photons exciting a quantum three-level particle reproduced most tension transient curves of water bug $Lethocerus \;Maximus$.

\end{abstract}

\pacs{}

\maketitle

\tableofcontents

\section{Introduction}

\begin{figure}[htbp]
\centering
\par
\begin{center}
\includegraphics[width=0.48\textwidth]{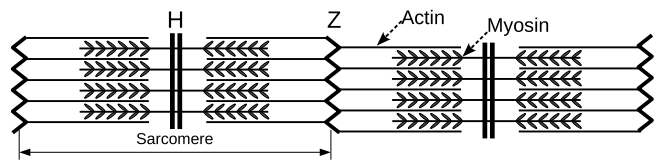}
\end{center}\vspace{-0.3cm}
\caption{\label{muscle} The longitudinal section of filament array in muscle fibre. Myosin molecules are long polymers as cross-bridges between the actin filament and myosin filaments.} \vspace{-0.2cm}
\end{figure}

Muscle is a bundle of cylindrical muscle fibres. One muscle fibre enclose hundreds of thinner cylindrical fibre called fibril. 
A fibril is a train of sarcommeres connected head to tail. The sarcomere, as seen by light microscopy, is a composite structure of actin filament and myosin filament(Fig. \ref{muscle}). The two filaments are bridged by long myosin protein called myosin molecule motor(see Ref. \cite{Aidley}\cite{Bagshaw} or biology textbook for a more detail review).

The length of a sarcomere is observed to decrease during muscle contraction. Huxley modeled the muscle contraction as mutual movement between actin filament and myosin filament\cite{Huxley}. When the myosin molecule in the myosin filament is excited up to attach on the actin filament, myosin molecule would combine with Adenosine Triphosphate molecules, and convert energy from Adenosine Triphosphate hydrolysis into mechanical force, this is the attached state. In the detached state, myosin molecule is at rest and doing nothing. 
The arrangement of myosin molecules along the length of filament is incommensurate. A stochastic model was introduced to describe the cooperative behavior of molecules motors which has disordered arrangement along a backbone\cite{Julicher}. Experiment research is beyond the description of classical mechanics two-state cross-bridge model\cite{Zahalak} which assumed the probability of the two states---detached state and attached state---satisfy a pair of coupled differential equations.

Myosin molecules looks like a long arm ended by a head domain. The myosin in muscle cell is usually called myosin-II, for it has two heavy chains in the head domain. A second kind of myosin molecule with single heavy chain was found in non-muscle cells, they are termed as myosin-I. The average length of myosin-II is about 160 nm\cite{Bagshaw}. Visible light corresponds to a wavelength range of 400-700 nm. The electromagnetic wavelength comparable to the length of myosin-II falls in ultraviolet region.

A muscle fibre surrounded by membrane is under control of nerve cell. Membrane is a filter system which is permeable for certain ions but is impermeable for other ions. The imbalance distribution of ions across the membrane results in an electric potential difference of -60 to -90 mv. The electric signals from nerve cell can modify the permeability of membrane. An active muscle would generate electric signal. Scientist use this electric signal to represent the tension inside the muscle. If we insert a needle with two fine-wire electrodes into the muscle, the electric activity of muscle can be detected and recorded by an oscilloscope. This electromyogram is in widespread use for medical examination of muscle.

Electromagnetic wave is termed as photon in quantum mechanics. The electric signal generated by active muscle is physically equivalent to a wave packet of photons. Quantum scattering between photons and molecules within biological system is a common phenomena, so does in muscle. Molecule only absorb photons at resonance frequency. We assume there exist a one-to-one correspondence between conformational change of motor molecule and the hopping between quantum states of molecule. 
We define the excited quantum level as the attached states in which the motor molecule is extended to a longe shape. The detached state is assumed to be equivalent to a quantum states with lower energy. The motor molecule becomes shorter in detached states. Besides the two quantum states of molecules, another quantum state of photon is introduced to induce the hopping between the two quantum level. Photon can propagate among different molecules. It is photon that 
couples different motor molecules to work together. Quantum physics has many techniques to control quantum states, thus it is possible to control the conformation of molecule motors by electromagnetic wave.

Muscle fibre can also be stimulated by chemical solutions which involves many complicate biochemistry\cite{Bagshaw}. 
There was an argument on chemical reactions in muscle using quantum mechanics\cite{McClare}, it is not related to 
my mathematical physics model. The time scale in my modeling is much larger than the chemical reaction. We only focus 
on the electrical stimulation of muscle fibre without adding any external chemical solutions. To make this mathematical physics model 
work for a real muscle fibre system, the following assumptions are necessary: 
(1) The strength of electric field should be so strong that it dominates the conformation of molecules over thermodynamic fluctuations. (2) 
For the muscle fibre without any external input of stimulus chemical ions, the conformational change of motor molecules governed by 
electromagnetic field is much larger than that induced by chemistry. This is because electric field is distributed across the whole space, 
it exerts a global force on the motor molecules. While chemical ion only interact with molecule locally, it does not come into action 
unless it finds the correct binding site. (3) The spatial conformation of motor molecule can be modified by concentration of some chemical ions.

We simplified the muscle fibre as a one dimensional chain of many giant quantum particles which represents molecule motors. 
The velocity of sliding actin filament acts as environmental parameter. These giant quantum particles are excited by absorbing photons, and relaxed by emitting photons. We arrange these particles regularly along a one dimensional chain. The wavelength of the stimulus electromagnetic pulse is assumed to be 
much larger than a sarcomere. The configuration of spatial arrangement within a sarcomere is supposed to has no influence on the output physics.

The article is organized as follows: 

In section II, we present the quantum Hamiltonian for deriving analogous force-velocity relation with respect to the coresponding macroscopic quantity. The empirical Hill's relation is consistent with the steady solution of Heisenberg equations. We predict the force velocity relation for the slow release case and 
the non-steady state solution. 

In section III, we study how the force decays after the stimulus of electric pulse. Exact solution of quantum two level model reproduced similar 
tension activation curves of cardiac muscle. The analytical solution of quantum three-level model can generate 
most similar curves of tension transients of insect flight muscle.

In section IV, the tension transients curve of skeletal muscle is mathematically reproduced from the point view of quantum coherent state. The experiment curve coincides with the third order projection of coherent state. 

In section V,  we derived the self-coupled quantum Hamiltonian for solving strongly coupled differential equations. For this case, 
the sliding velocity depends on the internal states of the motor molecule.

The last section is a brief summary and outlook.

\section{Force-velocity relation derived from quantum Hamiltonian}
\label{constvelocity}

In muscle release experiment, the two ends of the muscle is fixed at its resting length. Electric pulses with different frequency is superimposed to develop tension in the muscle. In a quick release experiment, one end of the muscle is set free quickly. For a slow release, the length of muscle is shortening with one end of the muscle oscillating or moving under control with in a relatively longer time interval. The tension-velocity relation is the relation between two macroscopic quantity. In the following, we shall express the force by microscopic density operator, while the velocity is taken as macroscopic parameter. We first assume the velocity of contraction is controlled by the density of some ions in the solution. We only model the excitation and decay process of muscle.

\begin{figure}[htbp]
\centering
\par
\begin{center}

\includegraphics[width=0.48\textwidth]{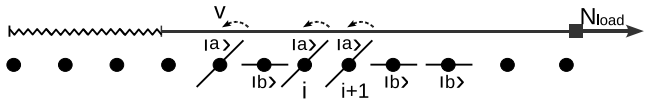}
\end{center}\vspace{-0.3cm}
\caption{\label{pair} A one dimensional chain of quantum two-level particles. Each particle represents one myosin molecule motor.} \vspace{-0.2cm}
\end{figure}

\subsection{Force-velocity relation for quick release of muscle}

We summarize a myosin molecule motor as a multi-level quantum particle. One eigenstate of the particle with respect to each level represents one conformation. In the excited state $|a\rangle$, myosin molecule is attached on actin filament. In the ground state $|b\rangle$, the quantum particle is at rest in myosin filament. One photon may excite the quantum particle from ground state to excited state. The particle hops from state $|a\rangle$ to state $|b\rangle$ by emitting one photon. The quantum representation of the mutual sliding movement between actin filament and myosin filament is to annihilate one quantum state on one lattice site and generate it again at the next nearest neighboring lattice site. As the sliding motion has only one direction, switching two neighboring lattice sites would contributes a minus sign in front of the hopping operator. The quantum Hamiltonian of this one dimensional chain(Fig. \ref{pair}) is unitary,
\begin{eqnarray}\label{Hamiltonian1}
H&=&\sum_{i}\omega_{i}^{a}|a\rangle_{i,i}\langle{a}|+\omega_{i}^{b}|b\rangle_{i,i}\langle{b}|+g_{ab}|a\rangle_{i,i}\langle{b}|
+g_{ba}|b\rangle_{i,i}\langle{a}|\nonumber\\
&+&\frac{iv}{s}[ |a\rangle_{i}\langle{a}|_{i+1}+|b\rangle_{i}\langle{b}|_{i+1}
-|a\rangle_{i+1}\langle{a}|_{i}-|b\rangle_{i+1}\langle{b}|_{i}].
\end{eqnarray}
where $s$ is the unit lattice spacing, and $v$ is the absolute sliding velocity. The hopping coefficient $g_{ab}(t)=p_{_{0}}E(t)$ is electric field. Usually we choose symmetric coupling coefficients, $g_{ab}(t)=g_{ba}(t)=g$. This Hamiltonian is neither single-particle Hamiltonian nor tight-binding model. For the particle at lattice site $i$, the probability of being in the excited level $|a\rangle_{i}$ is measured by the density operator, $|a\rangle_{i,i}\langle{a}|={\rho}_{aa}^i$. ${\rho}_{bb}^i=|b\rangle_{i,i}\langle{b}|$ defines the probability of being in state $|b\rangle_{i}$. ${\rho}_{ab}^i=|a\rangle_{i,i}\langle{b}|$ is proportional to the complex dipole moment. The probability density function are not independent, they must satisfy the constrain, ${\rho}_{aa}^i+{\rho}_{bb}^i=1$.

The time evolution of the density operators are governed by the equation of motion, $\dot{\rho}=-{i}{\hbar^{-1}}[H,\rho]$.
For the given Hamiltonian (\ref{Hamiltonian1}), the equations are 
\begin{eqnarray}\label{aa}
(\partial_{t}+{v}\partial_{x}){\rho}_{aa}^i&=&n^0_{a}+{i}g_{ab}{\rho}_{ab}^i{-i}g_{ba}{\rho}_{ba}^i-2\frac{v}{s}{\rho}_{aa}^i,
\nonumber\\
(\partial_{t}+{v}\partial_{x}){\rho}_{bb}^i&=&n^0_{b}+{i}g_{ba}{\rho}_{ba}^i{-i}g_{ab}{\rho}_{ab}^i-2\frac{v}{s}{\rho}_{bb}^i,
\end{eqnarray}

\begin{equation}\label{ab}
(\partial_{t}+{v}\partial_{x}){\rho}_{ab}^i=
{i}g_{ab}({\rho}_{bb}^i-{\rho}_{aa}^i)
{-i}\Delta\omega_{i}{\rho}_{ab}^i
-2\frac{v}{s}{\rho}_{ab}^i,
\end{equation}
where $\Delta\omega_{i}=\omega_{i}^{a}-\omega_{i}^{b}$ is the energy difference between the two states. At time $t_0$, and position $z_0$, the initial number of molecule motors in state $\mid{a}\rangle$ and $\mid{b}\rangle$ are denoted as $n^0_{a}$ and $n^0_{b}$ respectively, both $n^0_{a}$ and $n^0_{b}$ are functions of velocity $v$. In the steady states, the probability in both two sates are constants. Put the steady state constrain, $\partial_{t}{\rho}_{aa}^i=0$ and $\partial_{t}{\rho}_{bb}^i=0$, on the equations of motion---Eq. (\ref{aa}) and Eq. (\ref{ab}), one would arrive at an equality of probability distribution and velocity(detail calculation is presented in Appendix \ref{quickrelease}),
\begin{equation}\label{aa-v}
(\sum_{i}{\rho}_{aa}^i-\frac{N_{t}}{2})({v+sR(v)})=\frac{N_{t}s[n^0_{a}-n^0_{b}]}{4},
\end{equation}
where $R(v)$ is complicate function of velocity, 
\begin{eqnarray}
R(v)&=&-\frac{1}{8}
\frac{p^2_{_{0}}E^2_{n}}{\hbar^2}\frac{s}{2v}[G_{-}+G_{+}],\nonumber\\
G_{\pm}&=&\frac{4{v}^2/{s}^2}{4{v}^2/{s}^2+(\omega-f_{n}{\pm}kv)^2}.
\end{eqnarray}
where $k=\omega/c$, $c$ is the speed of light. $f_{n}$ is the frequency of the stimulus electric pulses. $N_{t}$ is the total number of particles. We define the tension of the one dimensional chain as the total number of particles in excited states, 
\begin{equation}\label{tension}
Tension=\chi\sum_{i}{\rho}_{aa}^i,
\end{equation}
where $\chi$ is a renormalization coefficient, we set it as $\chi=1$ for convenience. 
The myosin is attached on the actin filament in excited states. The more myosin molecules are excited, the stronger physical bonds there would exist between actin filament and myosin filament. The maximal tension is determined by the total number of particles, $Tension_{max}=N_{t}$, i.e., all particles are excited to connect the two filaments.

The existence of a maximal tension value is consistent with experiment measurement. If one apply a single electric shock across membrane, the tension of the muscle fibre first increase from zero to a transient maximal point and then decays to zero. The electric shock is an energy packet with finite number of photons. Suppose the energy packet contains $N_p$ photons, and $N_p<N_{t}$. It can maximally excite $N_p$ molecule motors. In the meantime, the motor is decaying and emitting photons. If $N_{t}$ photons are sent into the muscle at one time before any decay process, all the molecules will be excited, then it will reach the maximal tension state. It always take certain time for the muscle to eat all the photons, more photons must be added to fill the loss of photon by decay.

\begin{figure}[htbp]
\centering
\par
\begin{center}
\includegraphics[width=0.3\textwidth]{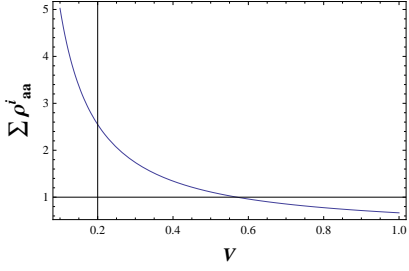}
\end{center}\vspace{-0.3cm}
\caption{\label{fv} The numerical plot of tension-velocity relation according to Eq. (\ref{aa-v}). } \vspace{-0.2cm}
\end{figure}

The relation between total probability density and velocity obtained for steady states, Eq. (\ref{aa-v}), is quite similar to  Hill's empirical force-velocity relation\cite{Hill}. Here we know the physical origin of every parameter. Although there is a complex function $R(v)$ involved in Eq. (\ref{aa-v}), the tension-velocity plots a hyperbolic curve, tension decreases with increased velocity(Fig. \ref{fv}). The equations of motion, Eq. (\ref{aa}) and Eq. (\ref{ab}), suggest that the velocity is actually a decay coefficient. Larger velocity leads to faster decay. Thus the number of physical bonds connecting the two filament decreases as velocity increases. This give us a natural explanation to experimental results.

\subsection{The equation of motion for the slow release}

Besides the decay induced by mutual sliding movement, there exist spontaneous decay due to thermal fluctuation and vacuum fluctuation. When we take the spontaneous decay into account, the equation of motion for the density operator is given by 
\begin{eqnarray}
\dot{\rho}&=& -\frac{i}{\hbar}[H,\rho]-\frac{1}{2}\left\lbrace \Gamma,\rho\right\rbrace,
\end{eqnarray}
where the $\Gamma$ is friction tensor, $\Gamma_{ij}=\gamma_{i}\delta_{ij}.$ The equation of motion has similar form as that for quick release, 

\begin{eqnarray}
(\partial_{t}+{v}\partial_{x}){\rho}_{aa}^i&=&n^0_{a}+{i}g_{ab}{\rho}_{ab}^i{-i}g_{ba}{\rho}_{ba}^i-\gamma'_{a}{\rho}_{aa}^i,
\nonumber\\
(\partial_{t}+{v}\partial_{x}){\rho}_{bb}^i&=&n^0_{b}+{i}g_{ba}{\rho}_{ba}^i{-i}g_{ab}{\rho}_{ab}^i-
\gamma'_{b}{\rho}_{bb}^i,
\end{eqnarray}

\begin{equation}
(\partial_{t}+{v}\partial_{x}){\rho}_{ab}^i=
{i}g_{ab}({\rho}_{bb}^i-{\rho}_{aa}^i){-i}\Delta\omega_{i}{\rho}_{ab}^i-\gamma'_{ab}{\rho}_{ab}^i,
\end{equation}
where $\Delta\omega_{i}=\omega_{i}^{a}-\omega_{i}^{b}$, $\gamma'_{ab}=\frac{1}{2}(\gamma'_{a}+\gamma'_{b})$. The decay coefficients of velocity now becomes the composition of velocity and spontaneous decay coefficients,  
\begin{equation}\label{gamma2}
\gamma'_{a}=(2\frac{v}{s}+\gamma_{a}),\;\;\;\;\gamma'_{b}=(2\frac{v}{s}+\gamma_{b}).
\end{equation}
For the steady state, $\partial_{t}{\rho}_{aa}^i=0$ ,   $\partial_{t}{\rho}_{bb}^i=0$, the relation equation between the excited state density operator and the composite decay coefficients is
\begin{eqnarray}\label{fv2}
(\sum_{i}{\rho}_{aa}^i-\frac{N_{t}}{2})\left[ \frac{\gamma'_{a}\gamma'_{b}}{\gamma'_{a}+\gamma'_{b}}+R(v)\right] =\frac{{\Delta}n(z)}{2}\frac{N_{t}\gamma'_{a}\gamma'_{b}}{\gamma'_{a}+\gamma'_{b}},
\end{eqnarray}
where ${\Delta}n(z,v,t)={n^0_{a}}{\gamma'_{a}}^{-1}-{n^0_{b}}{\gamma'_{b}}^{-1}$ is initial probability difference. $R(v)$ is a complex function of velocity, 
\begin{eqnarray}\label{slowR(v)}
R(v)&=&-\frac{1}{8}\frac{p^2_{_{0}}E^2_{n}}{\hbar^2}\gamma'^{-1}_{ab}[G_{-}+G_{+}],\nonumber\\
G_{\pm}&=&\frac{\gamma'^{2}_{ab}}{\gamma'^{2}_{ab}+(\omega-f_{n}{\pm}kv)^2}.
\end{eqnarray}
If the velocity term is much larger than spontaneous decay term, i.e., $2{v}/{s}\gg\gamma_{a},\gamma_{b}$, one neglects spontaneous decay, then Eq. (\ref{fv2}) reduced to former tension-velocity relation Eq. (\ref{aa-v}). Therefore,  Eq. (\ref{aa-v}) is the force-velocity relation for the quick release of muscle. If the muscle is released slowly, the spontaneous decay may become equally important as velocity term, $2{v}/{s}\approx\gamma_{a},\gamma_{b}$. In that case, the tension-velocity relation should obey Eq. (\ref{fv2}). Thus Eq. (\ref{fv2}) is the general force-velocity relation for steady state. As for the non-steady state, we have to solve the equations of motion to find how the excited state density operator evolves as a function of velocity.

\subsection{The tension-velocity relation beyond steady state}

Insect flight muscle oscillate far more rapidly than the frequency of the input nervous impulse. A muscle is unable to expand, it only contracts. The muscle expansion is achieved by contracting an opposing muscle. The oscillation frequency of muscle is the frequency of the oscillation between excited state and relaxed state. The $\omega_{i}^{a}|a\rangle_{i,i}\langle{a}|+\omega_{i}^{b}|b\rangle_{i,i}\langle{b}|$ terms in Hamiltonian Eq. (\ref{Hamiltonian1}) describes two free harmonic oscillator which are oscillating with frequency $\omega^{a}$ and $\omega^{b}$ respectively. The ultimate output frequency of Hamiltonian Eq. (\ref{Hamiltonian1}) is regulated by the hopping term between the two levels and the mutual sliding term. We perform Fourier transformation on the density operator, 
\begin{eqnarray}\label{al-ak}
\langle{a}_{l}|=\frac{1}{\sqrt{N}}\sum_{k}e^{iksl}\langle{a}_{k}|,
\;\;|a_{l}\rangle=\frac{1}{\sqrt{N}}\sum_{k}e^{-iksl}|a_{k}\rangle,
\end{eqnarray}
and substitute Eq. (\ref{al-ak}) into Hamiltonian Eq. (\ref{Hamiltonian1}). The Hamiltonian now becomes more brief, 
\begin{eqnarray}\label{Hkk}
H=H_{0}+\sum_{k}(g_{ab}|a_{k}\rangle\langle{b}_{k}|+g_{ba}|b_{k}\rangle\langle{a}_{k}|),
\end{eqnarray}
where $H_{0}$ is the exactly solvable part for regulated free oscillator,
\begin{eqnarray}\label{H0}
H_{0}=&\sum_{k} &\{[\omega^{a}-\frac{2v}{s}\sin(ks)]|a_{k}\rangle\langle{a}_{k}|\nonumber\\
&+&[\omega^{b}-\frac{2v}{s}\sin(ks)]|b_{k}\rangle\langle{b}_{k}|\}.
\end{eqnarray}
The regulated oscillation frequency depends on velocity and momentum vector $k$. The free oscillation modes has a maximal frequency and a minimum frequency, 
\begin{equation}
\left(\omega^{a}-\frac{2v}{s}\right)\leqq\left(\omega^{a}-\frac{2v}{s}\sin(ks)\right)
\leqq\left(\omega^{a}+\frac{2v}{s}\right).
\end{equation} 
The band width for existed free oscillation modes is determined by the sliding velocity. An oscillating electric field,
\begin{equation}
g_{ab}(t)=p_{_{0}}E(t)=p_{_{0}}E_{{0}}\cos({f}t),
\end{equation}
would induced the hopping between two levels. The state vector for the Schr$\ddot{o}$dinger wave equation, $\partial_{t}|\psi(t)\rangle=-i\hbar^{-1}H|\psi(t)\rangle$, is written as
\begin{eqnarray}
|\psi_{k}(t)\rangle&=&C_{a}\exp[-i(\omega_{k}^{a}-\frac{2v}{s}\sin(ks))t]\arrowvert{a}_{k}\rangle\nonumber\\
&+&C_{b}\exp[-i(\omega_{k}^{b}-\frac{2v}{s}\sin(ks))t]\arrowvert{b}_{k}\rangle,
\end{eqnarray}
The density operator are given by $\rho^{k}_{aa}(t)=C^{\ast}_{a}C_{a}$. Within rotating wave approximation, the terms that do not satisfy energy conservation are dropped, such as $d^{+}|a\rangle_{i,i}\langle{b}|$ and $d|b\rangle_{i,i}\langle{a}|$. The eigenstate density operator under the first order perturbation of the hopping terms $g_{ab}(t)$ is  
\begin{equation}\label{rhok}
\rho^{k}_{aa}=\frac{p^2_{_{0}}E^2_{0}}{4\hbar^2}\frac{\sin^{2}[\frac{1}{2}(\Delta\omega-({2v}/{s})\sin(ks)-{f})t]}
{[\Delta\omega-({2v}/{s})\sin(ks)-{f}]^2},
\end{equation}
where $\Delta\omega=\omega^{a}-\omega^{b}.$ The non-steady state tension decrease following $v^{-2}$ regulated by an oscillation factor. The ultimate oscillation frequency of tension includes three sources,
\begin{equation}\label{dispersion}
\omega_k=\Delta\omega-({2v}/{s})\sin(ks)-{f}.
\end{equation}
When the frequency difference of the two level vanishes, $\Delta\omega=0$, the output frequency is utterly dependent on the sum of perturbation frequency and sliding velocity. The fast oscillating ion flow of $K^+$, $Na^+$ across the membrane cause the potential to rise to +40 mv and restore its original value within a few milliseconds\cite{Bagshaw}. This oscillation frequency is coincide with some muscle oscillation. Some insect flight muscle contract 1000 times a second. My conjecture is that the oscillation frequency of muscle has some possible relation with the oscillating of ion flow. Certain insect flight muscle can oscillate more rapidly than the frequency of the input nervous impulse. Eq. (\ref{dispersion}) told us the output frequency not only depends on the frequency of input nervous impulse, but also depends on the sliding velocity and the frequency difference between the two levels. The sliding velocity may increase the output oscillation frequency. The output tension measured in experiment is the superposition of different oscillator modes, 
\begin{equation}
Tension=\sum_{k}\rho^{k}_{aa}(t). 
\end{equation}
The two ends of the one dimensional muscle fibre is always attached to some tissues. If there are $N+1$ particles along the chain, the beginning and the ending particle are fixed. As $\sin(ks)$ in dispersion Eq. (\ref{dispersion}) is a periodical function, independent standing wave modes is confined in the domain $(-\pi<ks<\pi)$. The allowed wave modes are
\begin{equation}
k=\frac{\pi}{L}, \frac{2\pi}{L}, \frac{3\pi}{L},...,\frac{m\pi}{L},...,\frac{(N_t-1)\pi}{L},
\end{equation}
where $L$ is the length of one sarcomere. All particles are oscillating in a standing wave pattern with frequency $\omega$. For a selected modes $\frac{m\pi}{L}$, the phase difference between nearest neighboring particles is $\frac{m\pi}{L}s$. If the sliding velocity is zero, ${v}=0$, the standing wave modes vanished.

\section{Quantum theory of tension transients under stretch activation}

A stretch activation is to lengthen the muscle fibre by external force, or to release the fibre to certain length with in a extremely short time. The fibres were glued between two glass rods. One rod is attached to the anode pin of a force transducer. The length and tension signal are displayed on a double beam oscilloscope\cite{Steiger}.

Both the myosin filament and actin filament are long helical chains. The elementary structure to build myosin filament looks like golf club. They entwined around each other to produce a helix with diameter of about 15 nm. The actin filament is a double helix structure composed of globular protein. The local electric charge distribution along the myosin chain and actin chain are not homogeneous. A mutual sliding between the myosin and actin chain would inevitable modify the local potential configuration. The raising or lowering of local potential barrier induce the hopping of electrons among different energy level. Many photons are generated within very short time. The stretch activation by mutual sliding movement is equivalent to stimulating the muscle fibre with an electric shock pulse.

The time evolution of tension induced by a stretch activation includes two periods. In the first period, there is a rapid decay due to the extremely large sliding velocity. A large number of photons are generated but are not absorbed by molecule motors. In the second period, the two filament stopped mutual sliding. The accumulated photon began coming into action. This period is dominated by the spontaneous emission and absorption of photon.

The ideal definition for contraction velocity is  
\begin{equation}
v(t_{0})=\lim_{\Delta{t}\rightarrow0}\frac{L(t_0+\Delta{t})-L(t_0)}{{\Delta}t},
\end{equation} 
$L$ is the length of sarcomere. In fact, what we measure in experiment is the average velocity within ${\Delta}t$. 
At the moment of quick release, the length experienced a sudden change, 
\begin{equation}
\Delta{L}(t_{0})=\int\Delta{L}(t)\delta(t-t_{0})dt.
\end{equation} 
The sliding velocity at $t_0$ is almost infinity, $v(t_{0})\approx\infty$. While after the stretch, the length does not change anymore, the sliding velocity becomes zero. The Heaviside function provides us a good mathematical description to the length variation during a stretch.

In the microscopic phenomena, a stretch activation would induce a sudden change of the particle distribution on different energy levels. There exist an optimal distribution which acts as the minimal point of a potential configuration. Any deviation from this optimal point has a tendency to go back. The optimal particle distribution is different for different types of muscle fibre. The external load applied upon the muscle is actually a high potential barrier. The particles are trapped in a local minimal point to counterbalance the external force. As a muscle is only able to contract, the intrinsic potential without external load is asymmetric.

\subsection{Quantum two level model for stretch activation of cardiac muscle}

The stretch activation of cardiac muscle has been extensively studied in experiment\cite{Steiger}. Various different species of vertebrates were put into use, including Humming bird, frog, Guinea pig, rat, rabbit. The tension transients following length perturbation were recorded by force transducer. At the instant lengthening the muscle, the force jumps up to a maximal force. Then the force drops to a temporary minimum. After several period of small oscillation, it finally reach a new maximal. The length of the muscle were kept invariant for a few seconds, and was suddenly release to its original length. A sharp drop of force appears immediately following by a fast recovery. The force transients for stretch and release looks more or less like mirror-image(Fig. 5. in Ref. \cite{Steiger}).

\begin{figure}[htbp]
\centering
\par
\begin{center}
\includegraphics[width=0.46\textwidth]{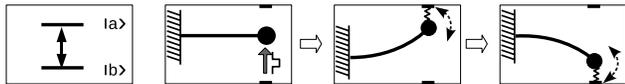}
\end{center}\vspace{-0.3cm}
\caption{\label{twolevel} The mechanical oscillator as an analogy of the quantum two level model.} \vspace{-0.2cm}
\end{figure}

A quantum two-level system provide a good model for the physics of force transients of cardiac muscle. A mechanical analogy of this quantum model is an oscillator in thick liquid. As shown in Fig. \ref{twolevel}, a heavy ball is attached to one end of a elastic rod. The sudden lengthening of muscle is to hit the ball with a pulse. The ball pops up to hit the upper level. Since the length of muscle after the sudden stretch keep invariant, a spring from the upper level will attach the ball to prevent it from going back. However the elastic rod will pull the ball down, the competition between the elastic rod and the spring from the upper level induced a local oscillation. When the kinetic energy is finally consumed by the friction of the thick liquid, the ball stopped oscillating. For the sudden release of muscle, the pule hit ball in the opposite direction of the lengthening. As the molecule motor runs only in one direction, it is reasonable to introduce a bias potential. This bias potential makes the ball a little bit harder to restore from the bottom level. Here the bias potential is gravitational field.

We assume the relaxed state of a muscle fibre is reached at an optimal ratio between attached states and detached states. 
One straightforward ratio is to set the number of particles in excited state and ground states as equal. This relaxed state is the vacuum state of the quantum two-level model. The vacuum state does not means there is no particle, but means the number of particle is exactly the number of antiparticle. In the mechanical analogy, this relaxed state corresponds to the equilibrium position of the ball without external pulses and gravitation field. If there exist bias field, the mechanical equilibrium is obtained by the balance between gravitational field and elastic force of the rod, $\kappa{\Delta{h}}=mg\Delta{h}$. We assume the optimal ratio is determined by the chemical potential of the two states, 
\begin{equation}
\rho^{0}_{aa}:\rho^{0}_{bb}=\exp[-\frac{\hbar\mu_a}{k_B{T}}]:\exp[-\frac{\hbar\mu_b}{k_B{T}}].
\end{equation}
where $\mu_a$ and $\mu_b$ are the chemical potential of excited state and ground state respectively. In the following, the density matrix is actually the difference between the usual density operator and the vacuum density operator, we simply denote  $({\rho}_{\alpha\beta}-\rho^{0}_{\alpha\beta})\rightarrow{\rho}_{\alpha\beta}$ for convenience. The vacuum density distribution determines the resting stiffness of a muscle fibre.

\begin{figure}[htbp]
\centering
\par
\begin{center}
$
\begin{array}{c@{\hspace{0.03in}}c}
\includegraphics[width=0.24\textwidth]{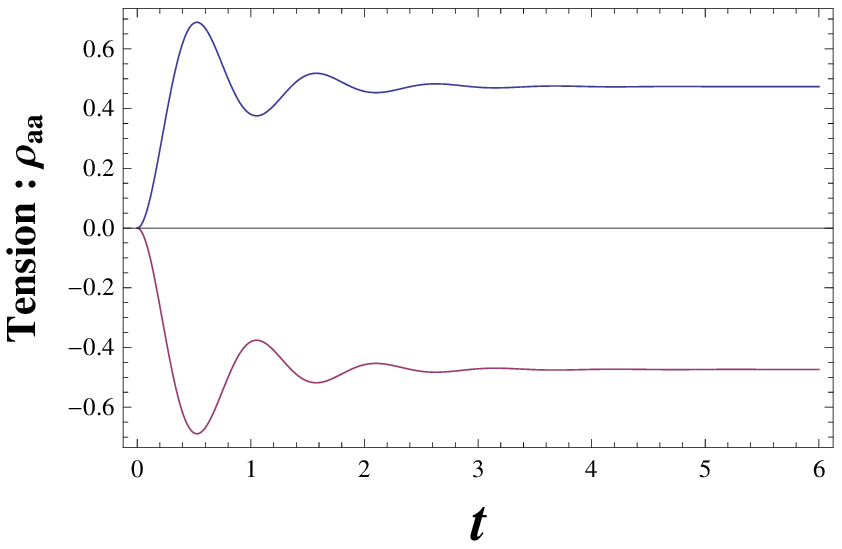}&\includegraphics[width=0.24\textwidth]{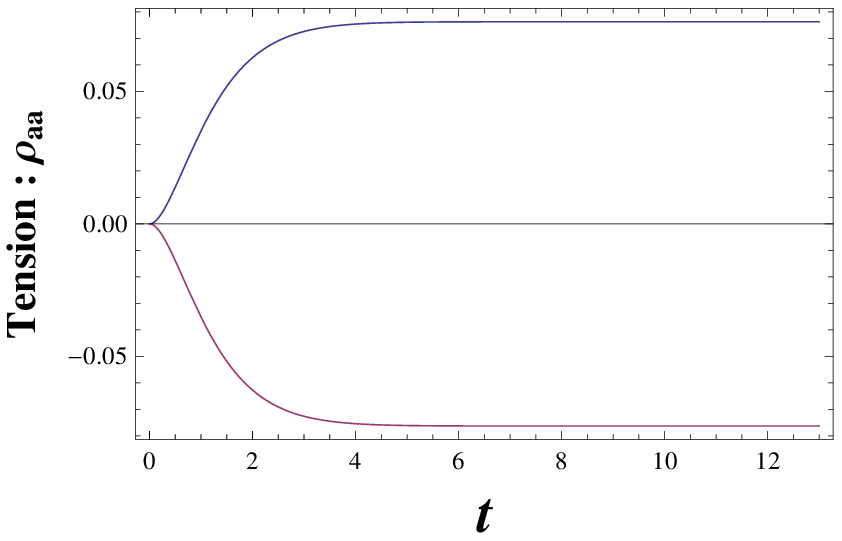}\\
\mbox{(a)} & \mbox{(b)}\\
\end{array}
$
\end{center}\vspace{-0.3cm}
\caption{\label{high-low} (a) The upper curve is the tension transient of ${\rho}_{aa}$ for a strong pumping field and low damping rate, $g/\gamma$=3. The lower curve is the time evolution of $({\rho}_{bb}-1)$. (b) The tension transition for a weak pumping field and high damping rate, $g/\gamma$=0.3. The upper curve is ${\rho}_{aa}(t)$, the lower is $({\rho}_{bb}-1)$.} \vspace{-0.2cm}
\end{figure}

\begin{figure}[htbp]
\centering
\par
\begin{center}
\includegraphics[width=0.3\textwidth]{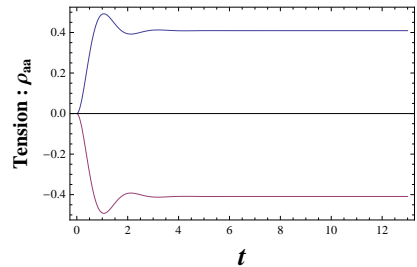}
\end{center}\vspace{-0.3cm}
\caption{\label{midfriction} The time evolution of ${\rho}_{aa}(t)$ for $g/\gamma$=1.5(the upper curve). The lower curve is $({\rho}_{bb}(t)-1)$ for $g/\gamma$=1.5.} \vspace{-0.2cm}
\end{figure}

A lengthening activation is equivalent to a positive pulse. All the particle in state $\arrowvert{b}\rangle$ are driven to state $\arrowvert{a}\rangle$.
\begin{eqnarray}
\Delta{L}(t_{0})=+N_{load},\;\;\;\rho_{aa}(t_{0})=1,\;\;\;\rho_{bb}(t_{0})=0.
\end{eqnarray} 
The sudden release at $t_1$ is negative pulse, all particles shift to state $\arrowvert{b}\rangle$.
\begin{eqnarray}
\Delta{L}(t_{1})=-N_{load},\;\;\;\rho_{aa}(t_{1})=0,\;\;\;\rho_{bb}(t_{1})=1.
\end{eqnarray} 
Thus the length perturbation provide the initial condition for time evolution of density operator. Since the length is constant after the stretch activation, we ignore the velocity term in Hamiltonian Eq. (\ref{Hamiltonian1}). The equation of motion including spontaneous decay reads,
\begin{eqnarray}\label{de}
\partial_{t}{\rho}_{aa}&=&{i}g_{ab}{\rho}_{ab}{-i}g_{ba}{\rho}_{ba}-\gamma_{1}{\rho}_{aa},\nonumber\\
\partial_{t}{\rho}_{bb}&=&{i}g_{ba}{\rho}_{ba}{-i}g_{ab}{\rho}_{ab}+\gamma_{1}{\rho}_{aa},\nonumber\\
\partial_{t}{\rho}_{ab}&=&{i}g_{ab}({\rho}_{bb}-{\rho}_{aa}){-i}\Delta\omega{\rho}_{ab}
-\gamma_{2}{\rho}_{ab},\nonumber\\
\partial_{t}{\rho}_{ba}&=&-{i}g_{ab}({\rho}_{bb}-{\rho}_{aa}){+i}\Delta\omega{\rho}_{ba}
-\gamma_{2}{\rho}_{ba},
\end{eqnarray}
where $\Delta\omega=\omega^{a}-\omega^{b}$. The coupling coefficient are chosen as the symmetric, $g_{ab}=g_{ba}=g.$ We focus on the case of two degenerated states, i.e., $\Delta\omega=0.$ In the beginning, all particles are in state $\arrowvert{b}\rangle$,  $\rho_{_{\alpha\beta}}(0)=\delta_{\alpha{b}}\delta_{{\beta}b}.$ This corresponds to a quick release. Applying the standard ansatz
\begin{equation}
\rho_{_{\alpha\beta}}(t)=\sum_{q=1}^{4}c_{\alpha\beta}^{q}\exp[\lambda_{q}t],\;\;\;\;\alpha,\beta=a,b,
\end{equation} 
one reduces the differential equation to algebra equation\cite{Vogel}. It finally yields the time evolution of probability in state $\arrowvert{a}\rangle$\cite{Vogel}, 
\begin{equation}\label{taa}
\rho_{aa}(t)=\frac{2g^2}{4g^2+\gamma_{1}\gamma_{2}}\left[1+\frac{\lambda_{1}e^{\lambda_{1}t}}{\lambda_{1}-\lambda_{2}}
+\frac{\lambda_{1}e^{\lambda_{2}t}}{\lambda_{2}-\lambda_{1}}\right].
\end{equation}
This is the conventional solution in the textbook of quantum optics\cite{Vogel}. The plot of this solution for three different cases were also shown 
in the book of Ref.\cite{Vogel}. I will reproduce some figures of Ref.\cite{Vogel} in the following to show what the tension transients curve looks like. 
I will compare different cases with experimental curves which I have no copyright to use here. $\rho_{aa}(t)$ governs the tension transient. Here
\begin{equation}
\lambda_{1,2}=-\frac{1}{2}(\gamma_{1}+\gamma_{2})\pm\frac{1}{2}\sqrt{(\gamma_{1}-\gamma_{2})^2-16g^2}.
\end{equation} 
After the stretch activation, there is no more external stimulus. The molecules undergoes radiative damping through spontaneous emission. The time evolution of $\rho_{aa}(t)$ relies on the ratio between driven field $g$ and damping rate $\gamma$, here we have set $\gamma_{1}=2\gamma,$  $\gamma_{2}=\gamma$. If the driven field is much larger than damping field, there is an oscillation upon the exponential increase of tension(the upper curve in Fig. \ref{high-low} (a)). In the opposite case, the damping dominates the dynamic evolution, it is an exponential increasing without oscillation(the upper curve in Fig. \ref{high-low} (b)). As the density operator must satisfy ${\rho}_{bb}+{\rho}_{aa}=1$, the time evolution of ${\rho}_{bb}=1-{\rho}_{aa}$ just follow the image curve of ${\rho}_{aa}$. For the intermediate case, the pumping strength is comparable with damping rate, both the lifetime and amplitude of oscillation are reduced, it approaches to an exponential increase for ${\rho}_{aa}$(the upper curve in Fig. \ref{midfriction}) or an exponential decay for ${\rho}_{bb}$(the lower curve in Fig. \ref{midfriction}).

Comparing these tension transients curve with the experiment records in Ref. \cite{Steiger}, we find ${\rho}_{aa}(t)$ in Fig. \ref{high-low} (a)) fits well with the muscle tension transient under stretch activation, including the tension-time course of rabbit papillary muscle in Fig. 2 of Ref. \cite{Steiger}, the tension-time course of cardiac papillary muscle and insect flight muscle in Fig. 26 of Ref. \cite{Steiger}. As for the release activation, 
the tension-time course of rabbit papillary muscle of Fig. 26. A of Ref. \cite{Steiger} and Fig. 2 of Ref. \cite{Steiger} fits very well with ${\rho}_{aa}(t)$ 
in the theoretical Fig. \ref{midfriction}.

The tension-time course of Humming bird muscle under stretch activation and release activation looks like mirror image, however the life time and amplitude of tension oscillation under stretch activation is larger than that for the release case(Fig. 5 of Ref. \cite{Steiger}). The stretch activation curve is well fitted by ${\rho}_{bb}(t)-1$ in Fig. \ref{high-low} (a)). The release activation curve fits better with ${\rho}_{aa}(t)$ in Fig. \ref{midfriction}. The release activation is the tension recovery from state $\arrowvert{b}\rangle$, the stretch activation is the recovery from $\arrowvert{a}\rangle$. The asymmetric tension-time course indicates the damping rate for $\arrowvert{b}\rangle$ is higher than the damping for state $\arrowvert{a}\rangle$. In the mechanical analogical phenomena, the damping may comes from friction of liquid or the bias external field. In polymer physics, the extension process of a long polymer is not always exactly 
the reverse process of contraction. I guess the motor molecule has similar behavior, this leads to the 
The asymmetric tension time course. We may call this a hysteresis phenomena.

The tensions time course for a weak pumping field and high damping rate(Fig. \ref{high-low} (b)), $g/\gamma$=0.3, reproduced the high tension state of rabbit psoas muscle. The tension transient of stretch and release are exactly mirror-image of exponential decay(Fig. 15 A of Ref. \cite{Steiger}).

The tension transient Eq. (\ref{taa}) has two extremal case, one is for extremely high friction $\gamma\gg{g}$, the tension first decreases, and then exponentially rise to a stable value. The other case is for very strong pumping field, ${g}\gg\gamma$. In this case, the tension shows swift oscillation, the envelop of the oscillation is an exponential decay curve. These two cases have no resemblance among the tension course of the experiment in Ref. \cite{Steiger}. There might exist some mechanism to prevent the cardiac muscle falling into these two extremal cases.

\begin{figure}[htbp]
\centering
\par
\begin{center}
$
\begin{array}{c@{\hspace{0.03in}}c}
\includegraphics[width=0.24\textwidth]{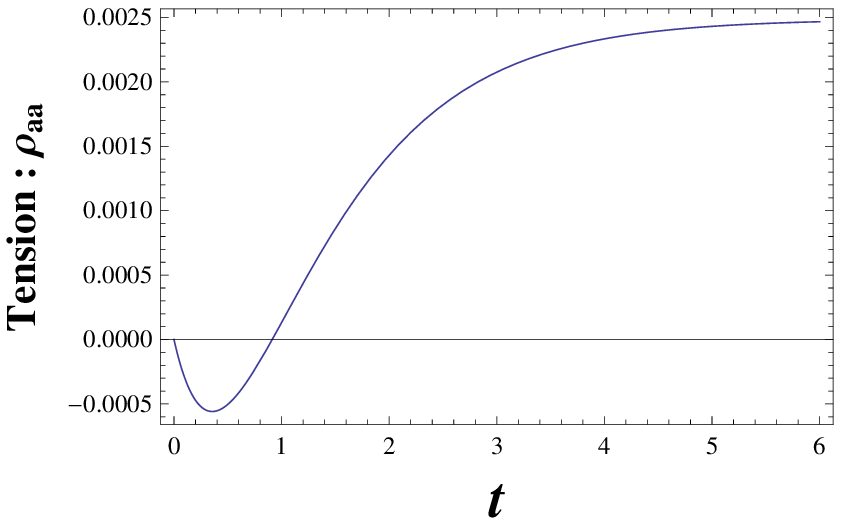}&\includegraphics[width=0.23\textwidth]{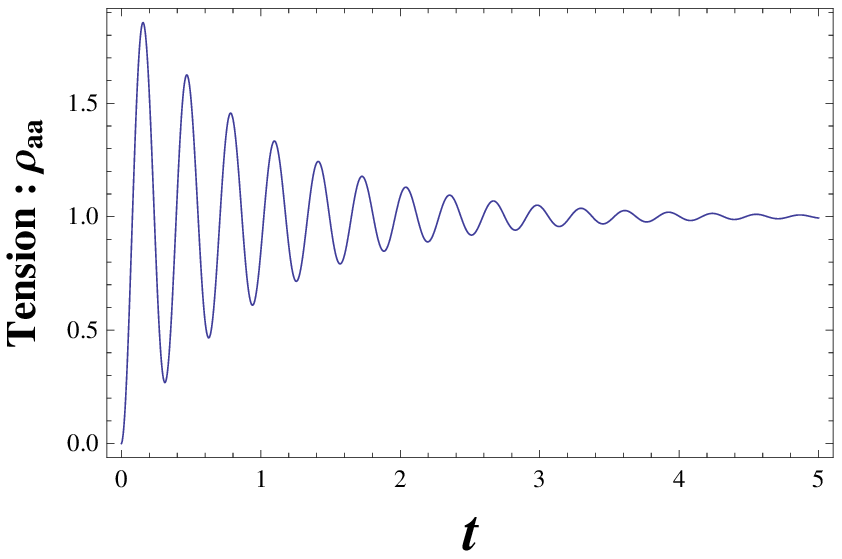}\\
\mbox{(a)} & \mbox{(b)}\\
\end{array}
$
\end{center}\vspace{-0.3cm}
\caption{\label{cos20} (a)The time evolution of ${\rho}_{aa}(t)$ for the case that friction $\gamma$ is much higher than pumping field $g$, $g/\gamma$=0.05.  (b) The time evolution of ${\rho}_{aa}(t)$ for $g/\gamma\gg1$, the driving field is much higher than friction.} \vspace{-0.2cm}
\end{figure}

\subsection{Quantum three-level model for tension transients in insect flight muscle}

The tension transients of insect flight muscle under length perturbation is different from that of vertebrate including rabbit, humming birds, frog, pig, and so on. The experiment in Ref. \cite{Steiger} reported the muscle tension course of water bug $Lethocerus \;Maximus$. If the muscle is prepared in a regular solution without ${Ga}^{++}$, the muscle tension almost has no response to length perturbation. In the presence of ${Ga}^{++}$ and  ${Mg}^{++}$, significant tension transient of $Lethocerus \;Maximus$ muscle were observed in reaction to length perturbation\cite{Steiger}. 

The tension course curve under both stretch and release activation showed a fast rise to a transient maximal and then followed by a slow decay\cite{Steiger}. The shape of the decaying tail is beyond the theoretical prediction of the quantum two level model. While a quantum three-level model fits the experiment data very well. This suggests that the molecule motors in insect flight muscle may be different from the motors in cardiac muscle.

We still model the insect flight muscle as one dimensional chain of quantum molecule motors. Now the molecule motor has three states: the attached state $\arrowvert{a}\rangle$,  the metastable state $\arrowvert{b}\rangle$ and the detached state $\arrowvert{c}\rangle$. We assume the hopping between different states is induced by vacuum fluctuation. Since muscle works at room temperature, the stochastic environmental fluctuation may destroy the coherent correlation between different motors. The motors are almost independent from each other, thus the one particle model is a good approximation. The usual Hamiltonian 
for a three-level particle interacting with vacuum modes reads, 
\begin{eqnarray}\label{H3}
&&H=i\sum_{k}[g^{k}_{_{1}}e^{i(\omega_{ac}-\omega_{k})t}d_{k}\arrowvert{a}\rangle\langle{c}\arrowvert
+g^{k}_{_{2}}e^{i(\omega_{bc}-\omega_{k})t}d_{k}\arrowvert{b}\rangle\langle{c}\arrowvert]\nonumber\\
&&-i\sum_{k}[g^{k}_{_{1}}e^{-i(\omega_{ac}-\omega_{k})t}d^{\dag}_{k}\arrowvert{c}\rangle\langle{a}\arrowvert
+g^{k}_{_{2}}e^{-i(\omega_{bc}-\omega_{k})t}d^{\dag}_{k}\arrowvert{c}\rangle\langle{b}\arrowvert],\nonumber\\
\end{eqnarray}
where $\omega_{\alpha\beta}=\omega_{\alpha}-\omega_{\beta}$ is the frequency difference between state $\arrowvert{\alpha}\rangle$ and $\arrowvert{\beta}\rangle$, $\alpha,\beta=a,b,c$. $d_{k}$ and $d^+_{k}$ are the annihilation and creation operator of the $k$th vacuum mode with frequency $\omega_{k}$. $g^{k}_{_{1}}$ and $g^{k}_{_{2}}$ denotes the coupling strength between vacuum modes and the three-level particle.

\begin{figure}[htbp]
\centering
\par
\begin{center}
\includegraphics[width=0.47\textwidth]{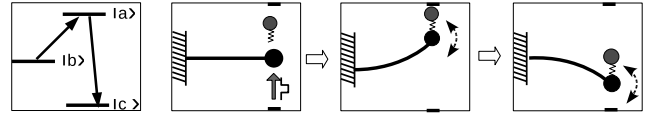}
\end{center}\vspace{-0.3cm}
\caption{\label{threelevel} The mechanical oscillator analogy of the quantum three level model.}\vspace{-0.2cm}
\end{figure}

The quantum three-level system has a similar mechanical analogy with the damping oscillator for the two-level model. The key difference is that the oscillator has a metastable partner suspended above the equilibrium position(Fig. \ref{threelevel}). 
To perform a lengthening perturbation, we sent a shock pulse to hit the oscillator in the opposite direction of gravitation. The oscillator will rise to attach the partner ball and becomes a pair. The pair first rise to the attached state$\arrowvert{a}\rangle$, and then decayed to the detached state $\arrowvert{c}\rangle$. For a sudden release perturbation, the shock pulse hits the ball along the direction of gravitation field. The oscillator first drops to the detached state $\arrowvert{c}\rangle$, and then is pulled up by the elastic rod. It continues to rise high enough to attach the metastable ball, and oscillates as a pair. This damping oscillator model gives us a rough phenomena of tension time course for insect flight muscle.

We assume the myosin molecule motors as an equivalent three-level system. The tension of the one dimensional chain is proportional to the total number of particles in the attached state, 
$Tension\propto\sum_{i}{\rho}_{aa}^i$. Another equivalent way of modeling this system is to take the whole chain as one giant three-level particle. The tension of the particle is proportional to the probability in the attached state, in this case, we take muscle as single particle system which is relatively easy to handle with. An analytical investigation of Hamiltonian Eq. (\ref{H3}) was well performed for the spontaneous emission of a three-level atom with quantum interference\cite{Zhu}. The dynamics of a quantum particle is governed by the Schr$\ddot{o}$dinger equation, $\partial_{t}|\psi(t)\rangle=-i\hbar^{-1}H|\psi(t)\rangle$. The general state vector is written as
\begin{equation}
\arrowvert{\psi(t)}\rangle=C_{a}\arrowvert{a}\rangle\arrowvert{0}\rangle+C_{b}\arrowvert{b}\rangle\arrowvert{0}\rangle+\sum_{k}C_{k}d^{\dag}_{k}\arrowvert{c}\rangle\arrowvert{0}\rangle.
\end{equation} 
The molecule is initially in the attached state $\arrowvert{a}\rangle$ and metastable state $\arrowvert{b}\rangle$. When the frequency difference between the attached state and metastable state is much smaller than the frequency difference between the attached state and detached state, i.e., $\omega_{ba}\ll\omega_{bc},\omega_{ac}$, the Schr$\ddot{o}$dinger equation admits an analytical solution\cite{Zhu}, 
\begin{eqnarray}\label{cacb}
C_{a}(t)&=&\frac{-2}{\sqrt{\gamma_{a}\gamma_{b}}}(A\lambda_{1}e^{\lambda_{1}t}+B\lambda_{2}e^{\lambda_{2}t})e^{-(\gamma_{b}/2+i\omega_{ba})t},\nonumber\\
C_{b}(t)&=&(Ae^{\lambda_{1}t}+Be^{\lambda_{2}t})e^{-(\gamma_{b}/2)t}.
\end{eqnarray} 
where the coefficients $A$ and $B$ are
\begin{eqnarray}
A&=&\frac{\lambda_{2}C_{b}(0)+0.5\sqrt{\gamma_{a}\gamma_{b}}C_{a}(0)}{\lambda_{2}-\lambda_{1}},\nonumber\\
B&=&\frac{\lambda_{1}C_{b}(0)+0.5\sqrt{\gamma_{a}\gamma_{b}}C_{a}(0)}{\lambda_{1}-\lambda_{2}}.
\end{eqnarray} 
The decay rate of the two upper level are $\gamma_{b}=2(\pi{g_{_{1}}})^2D(\omega_{b})$, $\gamma_{a}=2(\pi{g_{_{2}}})^2D(\omega_{a})$. $D(\omega)$ is the mode density. $\lambda_{1,2}$ determines the amplitude of the decaying metastable state, 
\begin{equation}
\lambda_{1,2}= \frac{\Delta\gamma}{4}+\frac{i\omega_{ba}}{2}\pm{\sqrt{(\frac{\Delta\gamma}{4}
+\frac{i\omega_{ba}}{2})^{2}+\frac{\gamma_{a}\gamma_{b}}{4}}},
\end{equation} 
where $\Delta\gamma=\gamma_{b}-\gamma_{a}$ is the difference of decay rates for the two upper level. The probability of being in the attached state is given by the density operator ${\rho}_{aa}(t)$. Here the weights of state vector, $C_{a}(t)$ and $C_{b}(t)$, define ${\rho}_{aa}(t)$ and ${\rho}_{bb}(t)$,
\begin{equation}
{\rho}_{aa}(t)=C^{\ast}_{a}(t)C_{a}(t),\;\;\;\;{\rho}_{bb}(t)=C^{\ast}_{b}(t)C_{b}(t).
\end{equation} 
This density operator is actually a reduced density operator by tracing out all vacuum modes, 
\begin{equation}
{\rho}_{\alpha\alpha}=Tr_{_{k}}(\arrowvert{\alpha},k\rangle\langle{\alpha},k\arrowvert),\;\;\;\;{\alpha}=a,b,c.
\end{equation}
To get a clear comparison with the two-level model, we map the Schr$\ddot{o}$dinger equations of motion for $C_{\alpha}$ \cite{Zhu} into the equations of density operator, 
\begin{eqnarray}\label{rho3}
\partial_{t}{\rho}_{aa}&=&-\gamma_{a}{\rho}_{aa}-\frac{\sqrt{\gamma_{a}\gamma_{b}}}{2}(e^{-i\omega_{ba}t}{\rho}_{ba}+e^{i\omega_{ba}t}
{\rho}_{ab}),\nonumber\\
\partial_{t}{\rho}_{bb}&=&-\gamma_{b}{\rho}_{bb}-\frac{\sqrt{\gamma_{a}\gamma_{b}}}{2}(e^{i\omega_{ba}t}{\rho}_{ab}
+e^{-i\omega_{ba}t}{\rho}_{ba}),\nonumber\\
\partial_{t}{\rho}_{ab}&=&-\frac{\gamma_{a}+\gamma_{b}}{2}{\rho}_{ab}
-\frac{\sqrt{\gamma_{a}\gamma_{b}}}{2}e^{-i\omega_{ba}t}({\rho}_{aa}+{\rho}_{bb}),\nonumber\\
\partial_{t}{\rho}_{ba}&=&-\frac{\gamma_{a}+\gamma_{b}}{2}{\rho}_{ba}
-\frac{\sqrt{\gamma_{a}\gamma_{b}}}{2}e^{i\omega_{ba}t}({\rho}_{aa}+{\rho}_{bb}).
\end{eqnarray}
The sum of the probability in the three levels satisfy the constrain, ${\rho}_{aa}+{\rho}_{bb}+{\rho}_{cc}=1$. For the two-level model, the sum of probability in the two level is conserved,  $\partial_{t}[{\rho}_{aa}+{\rho}_{bb}]=0$. While for the three-level model, ${\rho}_{aa}+{\rho}_{bb}=1-{\rho}_{cc}$, the sum of the probability in the upper two level depends on the probability in the detached state $\arrowvert{c}\rangle$. The equations of motion (\ref{rho3}) have already incorporated the evolution of ${\rho}_{cc}$. One may verify $\partial_{t}[{\rho}_{aa}+{\rho}_{bb}]\neq0$ in Eq. (\ref{rho3}).

The two excited states are coupled to the detached state by the same vacuum modes. The photon emitted by the transition from $\arrowvert{b}\rangle$ to $\arrowvert{c}\rangle$ may induce the hopping from $\arrowvert{c}\rangle$ to $\arrowvert{a}\rangle$. The photon may comes from one molecule, but absorbed by another molecule. Thus there exist strong quantum coherence between the two decay channel. A sudden lengthening or release of muscle leads to the mutual sliding between the charged filaments. A packet of photons are created immediately but are bot absorbed. The stretch activation and release activation set up the initial value of population at different level. The deviation from equilibrium position results in a target potential barrier, which keeps a certain number of excited molecule from tunneling into lower level. The number of survived attached states determine the final tension value after decay.

\begin{figure}[htbp]
\centering
\par
\begin{center}
$
\begin{array}{c@{\hspace{0.03in}}c}
\includegraphics[width=0.23\textwidth]{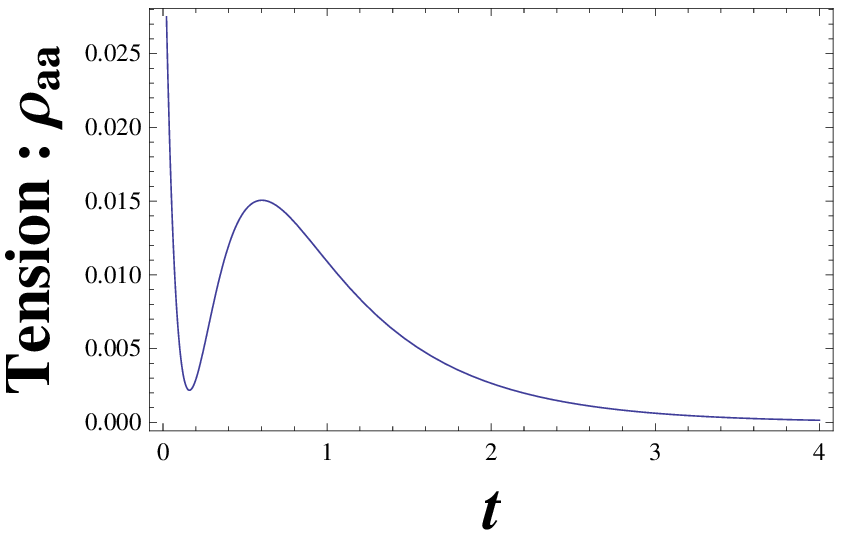}&\includegraphics[width=0.23\textwidth]{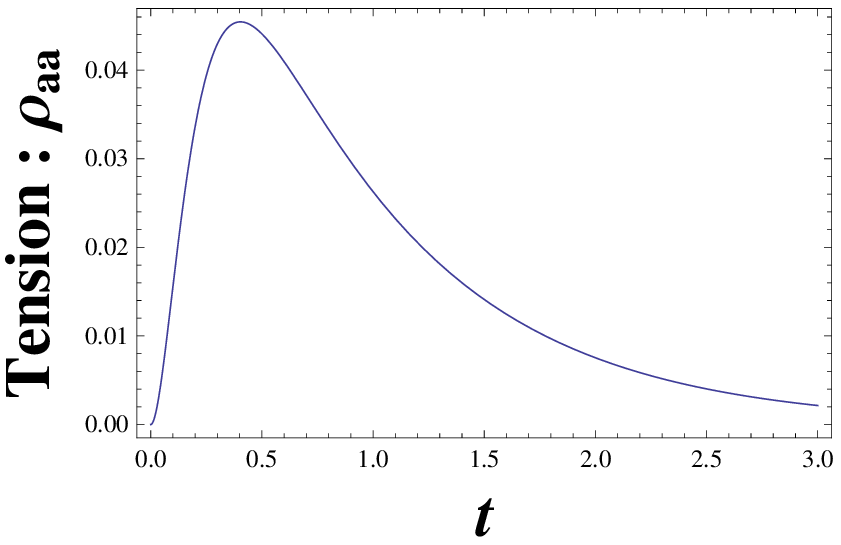}\\
\mbox{(a)} & \mbox{(b)}\\
\end{array}
$
\end{center}\vspace{-0.3cm}
\caption{\label{7-4-4} (a)Initial population in metastable state is ${\rho}_{bb}(0)=0.8$. The attached state is  ${\rho}_{aa}(0)=0.2$. Decay rate of metastable state is $\gamma_{b}=7$. Decay rate of attached state is $\gamma_{a}=4$. The frequency difference of the two excited states is $\omega_{ba}=4$. This curve corresponds to a stretch activation. (b) No population initially exist in the excited state ${\rho}_{aa}(0)=0$. ${\rho}_{bb}(0)=0.5$, $\gamma_{b}=7$, $\gamma_{a}=7$, $\omega_{ba}=4$. This corresponds to a release activation.} \vspace{-0.2cm}
\end{figure}

\begin{figure}[htbp]
\centering
\par
\begin{center}
$
\begin{array}{c@{\hspace{0.03in}}c}
\includegraphics[width=0.24\textwidth]{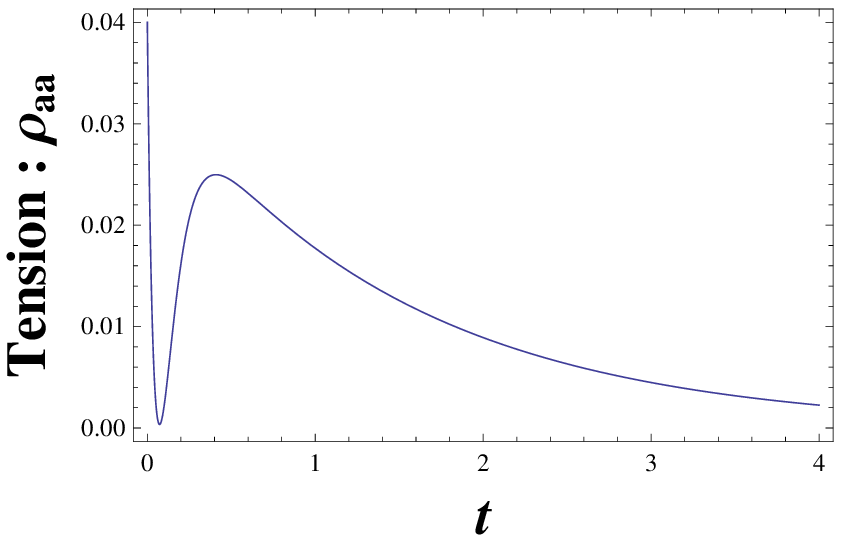}&\includegraphics[width=0.24\textwidth]{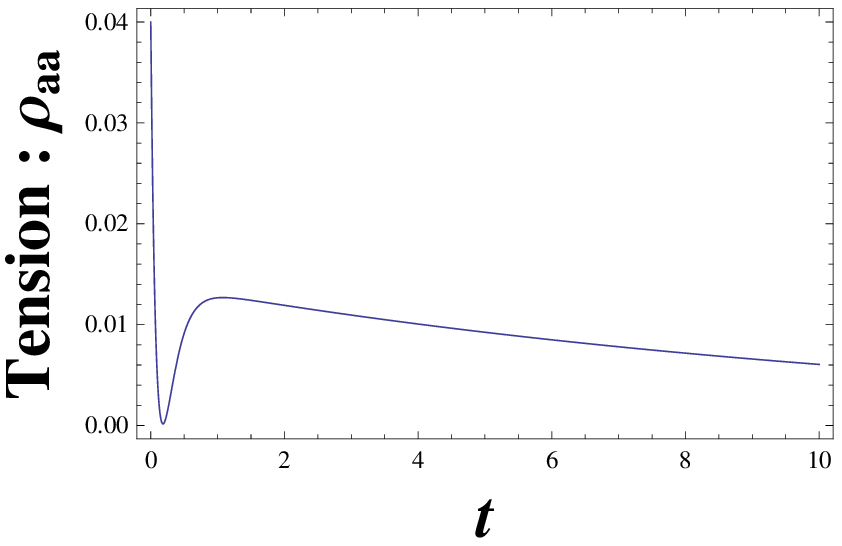}\\
\mbox{(a)} & \mbox{(b)}\\
\end{array}
$
\end{center}\vspace{-0.3cm}
\caption{\label{7-4-1} (a) Comparing with Fig. \ref{7-4-4}, the decay rates for attached state increases to $\gamma_{a}=14$.  $\gamma_{b}=7$,  $\omega_{ba}=4$, ${\rho}_{bb}(0)=0.5$, ${\rho}_{aa}(0)=0.2$. Tension decay becomes slower. (b) The decay rate $\gamma_{a}=4$. The frequency difference is reduced to $\omega_{ba}=1$. $\gamma_{b}=7$, ${\rho}_{bb}(0)=0.8$, ${\rho}_{aa}(0)=0.2$. } \vspace{-0.2cm}
\end{figure}

One qualitative conjecture of Ref. \cite{Steiger} is that ${Mg}^{++}$ is able to vary the ${Ga}^{++}$ sensitivity of the contractile system. When the $Lethocerus \;Maximus$ muscle is in a regular solution with only ${Ga}^{++}$ ions, the tension reaches the delayed maximum within in 70 msec\cite{Steiger}. When ${Mg}^{++}$ is added, the time to reach the maximum tension becomes much longer, it is about $500$ msec\cite{Steiger}. Fig. 18 B in Ref. \cite{Steiger} showed the tension time course in ${Ga}^{++}$ and  ${Mg}^{++}$ poor contraction solution. The experiment is performed at temperature T=$18^{\circ}C$ with 5 mM ATP. The ion concentration is estimated as pGa=8$\sim$9. pMg=7$\sim$8, pH=6.5. Another Fig. 23 in  Ref. \cite{Steiger} showed the case 
that the concentration of ${Ga}^{++}$ and ${Mg}^{++}$ is even much higher than that for Fig. 18 B in  Ref. \cite{Steiger}. 
The tension time course of Fig. 23 in  Ref. \cite{Steiger} is recorded at temperature T=$18.5^{\circ}C$ with 5 mM ATP, pH=6.5, pMg$\sim$5, pGa$\sim$6.8.

My analogue theoretical conjecture is the damping rate of exited states is related to the relative ratio of ${Mg}^{++}$ to ${Ga}^{++}$ and their absolute value of concentration. Increasing ${Mg}^{++}$ maybe reduce the decay rate. Higher concentration induces lower decay rate. This theoretical conjecture is based on 
the solution of three-level model in Eq. (\ref{cacb}). We found two very similar tension time course to the experimental curve of 
Fig. 18 B in Ref. \cite{Steiger}. The first Fig. \ref{7-4-4} (a) simulates the stretch activation, for which the initial probability in the attached state is ${\rho}_{aa}(0)=0.2$. The second Fig. \ref{7-4-4} (b) is for the release activation, no particle exists initially in the attached state. 
To get the similar tension transient curve with respect to Fig. 23 in  Ref. \cite{Steiger} from the three-level model, we raised the value of damping rate $\gamma_a$ from $4$ to $14$ in the numerical calculation. The theoretical output of Fig. \ref{7-4-1} can be identified as the experimental curve of Fig. 23 in  Ref. \cite{Steiger}. This theoretical curve in Fig. \ref{7-4-1} shows a slower decay process compared with the two decay curves lower concentration of ${Mg}^{++}$ and ${Ga}^{++}$(Fig. \ref{7-4-4}).

\begin{figure}[htbp]
\centering
\par
\begin{center}
$
\begin{array}{c@{\hspace{0.03in}}c}
\includegraphics[width=0.23\textwidth]{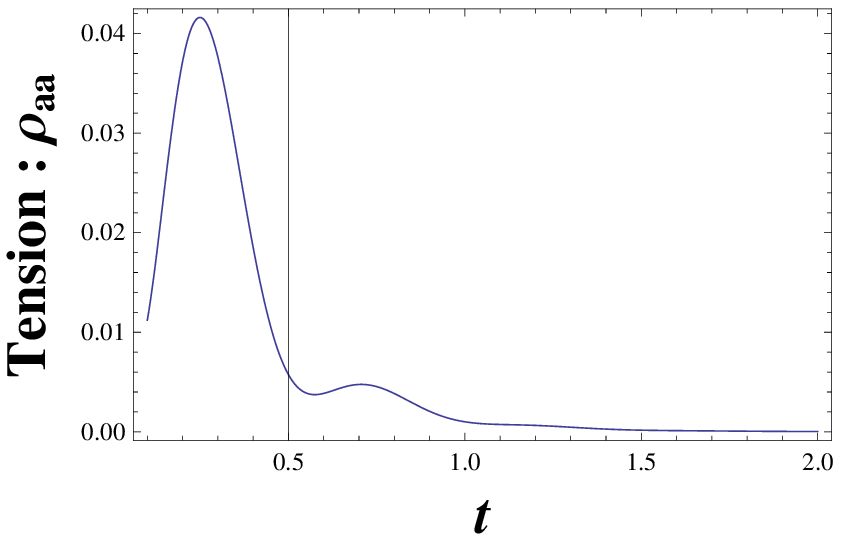}&\includegraphics[width=0.23\textwidth]{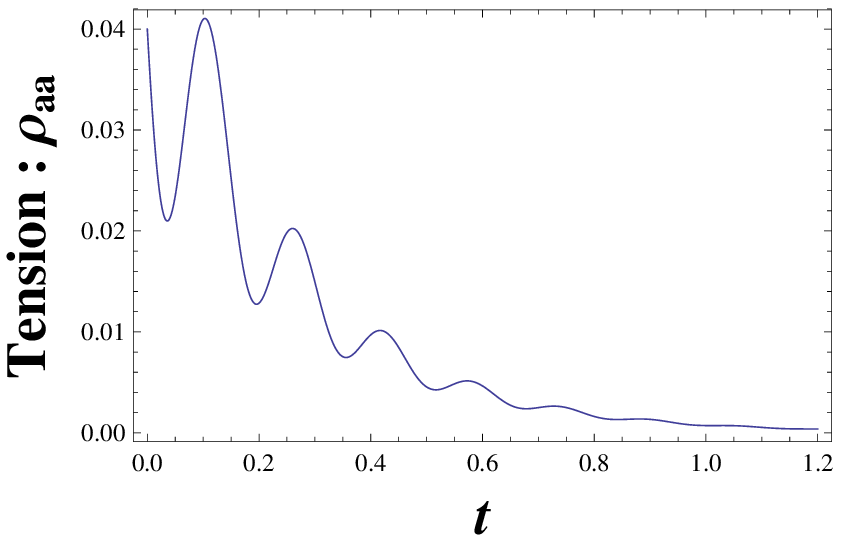}\\
\mbox{(a)} & \mbox{(b)}\\
\includegraphics[width=0.23\textwidth]{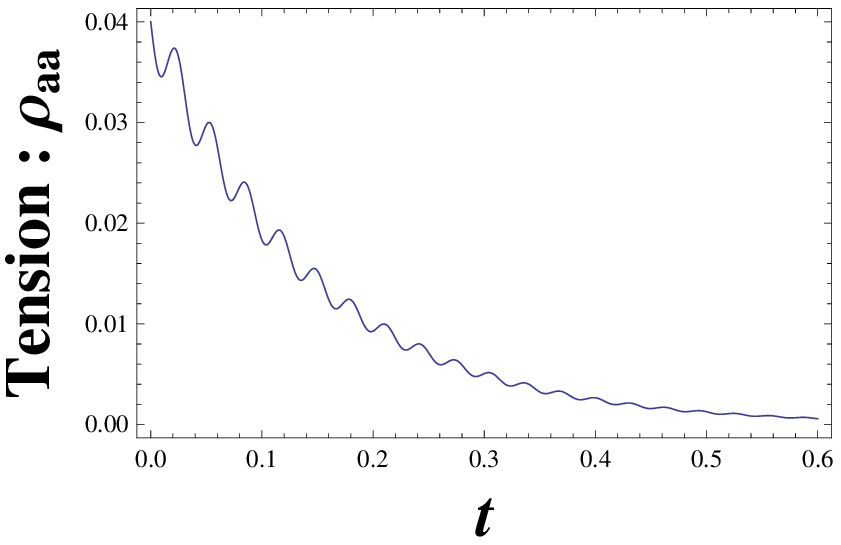}&\includegraphics[width=0.23\textwidth]{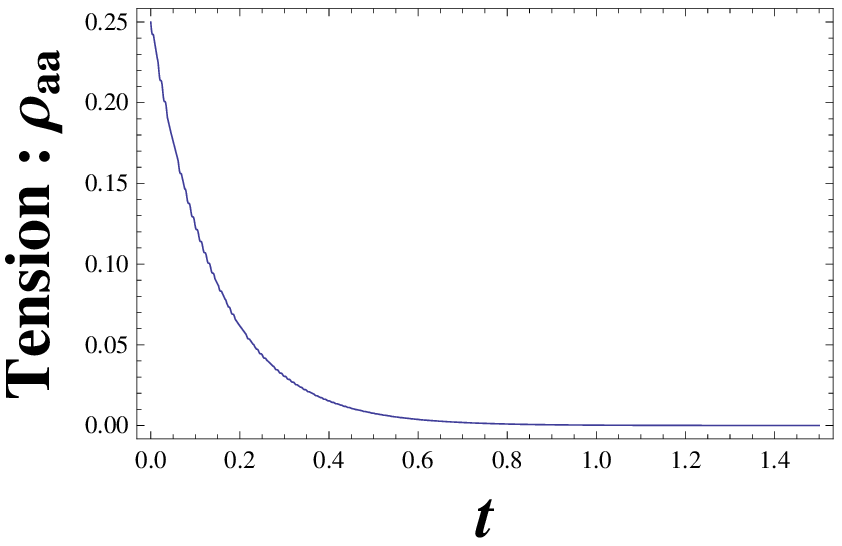}\\
\mbox{(c)} & \mbox{(d)}\\
\end{array}
$
\end{center}\vspace{-0.3cm}
\caption{\label{7-4-200} (a) The frequency difference between the metastable state and attached state $\omega_{ba}$ is very large, $\omega_{ba}=14$. ${\rho}_{bb}(0)=0.8$, ${\rho}_{aa}(0)=0.2$, $\gamma_{b}=7$, $\gamma_{a}=4$. (b) The frequency difference $\omega_{ba}$ increases to $\omega_{ba}=40$. ${\rho}_{bb}(0)=0.8$, ${\rho}_{aa}(0)=0.2$, $\gamma_{b}=7$, $\gamma_{a}=4$. (c) $\omega_{ba}$ is $\omega_{ba}=200$. ${\rho}_{bb}(0)=0.5$, ${\rho}_{aa}(0)=0.2$, $\gamma_{b}=7$, $\gamma_{a}=7$. (d) $\omega_{ba}$ is almost infinity, $\omega_{ba}=900$. ${\rho}_{bb}(0)=0.5$, ${\rho}_{aa}(0)=0.5$, $\gamma_{b}=7$, $\gamma_{a}=7$.} \vspace{-0.2cm}
\end{figure}

The frequency difference $\omega_{ba}$ indicates the oscillation between the two excited states. The exponential decay of tension is strongly regulated by this oscillation. For small value of $\omega_{ba}=14$, the tension has a dominant peak followed by small waves (Fig. \ref{7-4-200} (a)). If we magnify the tension course of experiment Fig. 19 in  Ref. \cite{Steiger}, the theoretical Fig. \ref{7-4-200} (a) is coincide with experimental curve. This curve of Fig. 19 in  Ref. \cite{Steiger} is obtained at temperature =$23^{\circ}C$ with 15 mM ATP, and pGa=8$\sim$9. pMg=7$\sim$8, pH=6.5. Comparing the experimental condition with that for Fig. 18 in Ref. \cite{Steiger}, one may see the temperature raise about $5^{\circ}C$, the concentration of ATP grows from 5 mM ATP to 15 mM ATP. My conjecture is that increasing ATP and temperature enhanced the oscillation between two excited states.

If we make the $\omega_{ba}$ larger, i.e., $\omega_{ba}=40$, the decay process shows more periods of oscillation(Fig. \ref{7-4-200} (b)). When $\omega_{ba}=200$, the tension-time course becomes a wavy curve of exponential decay(Fig. \ref{7-4-200} (c)). This wavy curve is quit similar to the tension transient of rabbit psoas muscle(Fig. 15 B in Ref. \cite{Steiger}). For an infinite large frequency difference, $\omega_{ba}=900$, the oscillation is completely suppressed. The tension transients approaches to a smooth exponential decay(Fig. \ref{7-4-200} (d)), which looks very similar to the experimental curve in the high tension states(Fig. 23. C, in  Ref. \cite{Steiger}). The high tension states of $Lethocerus \;Maximus$ muscle is reached by increasing pGa from 8$\sim$9 to 6.8, and keeping pMg=7$\sim$8. From the theoretical point of view, a large frequency difference $\omega_{ba}$ produced a large energy barrier. The quantum interference between the two excited states vanished. In mind of the theoretical approximation we used, $\omega_{ba}\ll\omega_{bc}$ and $\omega_{ba}\ll\omega_{ac}$, a large $\omega_{ba}$ indicates that the excited states are almost decoupled from ground state. The dynamics of the three-level system is dominated by decay of excited state. Since no pumping field can pass an infinite barrier, the evolution of excited states is an exponential decay.

When the frequency difference $\omega_{ba}$ vanishes, the two excited states becomes degenerated. Both the two states obey an exponential decay, 
\begin{eqnarray}
{\rho}_{aa}(t)&=&A^2\frac{\gamma_{b}}{\gamma_{a}}-ABe^{-t(\gamma_{a}+\gamma_{b})/2}+B^2\frac{\gamma_{a}}{\gamma_{b}}e^{-t(\gamma_{a}+\gamma_{b})},\nonumber\\
{\rho}_{bb}(t)&=&A^2+B^2e^{-t(\gamma_{a}+\gamma_{b})}+ABe^{-t(\gamma_{a}+\gamma_{b})/2}.
\end{eqnarray} 
The tension now is the sum of the probability of the two degenerated states, $Tension={\rho}_{aa}(t)+{\rho}_{bb}(t)$,
\begin{equation}\label{B2}
Tension=A^2(\frac{\gamma_{b}}{\gamma_{a}}+1)+B^2(\frac{\gamma_{a}}{\gamma_{b}}+1)e^{-t(\gamma_{a}+\gamma_{b})}.
\end{equation}
The tension evolution is an exponential decay plus a constant term. The constant tension depends on the initial value of the density operator. For this special case, three-level model with two degenerated excited states gives similar result as the two-level model. However, 
there is always an induced decay rate by emitting photons in three-level model due to the metastable state. 
In two-level model, the emitting photons can be brought back to excited state without violating selection rules, 
thus the output of two-level model does not produce all of the decay curves as the three-level model did.

\section{Projecting quantum coherent state for skeletal muscle fibre under electric stimulus}

Comparing with cardiac muscle and insect flight muscle, skeletal muscle has two special characters: one is low resting stiffness, the other is its ability of extending much beyond its optimal length. When we summarize these two special characters into theoretical modeling, low stiffness 
indicates there are less number of molecules motors in the attached state when the muscle is not excited. Two filaments connected by less bonding is easy to extend. The electrical signal representing muscle tension is physically explained as polarized photon. The strength of electric signal is proportional to the density of photon. Quantum coherent photons exist in many living biological system\cite{Popp}. Even though an isolated muscle fibre is not alive, we can still shed some new light on the twitch tension evolution in reaction to the stimulus of electric signal from the point of view of  coherent states.

External stimulus field will generate tension in the muscle. If the two ends of a muscle fibre is fixed, the tension developed by external stimulus is called isometric tension. The isometric tension of skeletal muscle was measured many decades ago\cite{Gabel}. A superamaximal rectangular pulses is used to stimulate the muscle. Both the second and third derivatives of twitch tension are recorded by some software\cite{Gabel}.

\begin{figure}[htbp]
\centering
\par
\begin{center}
\includegraphics[width=0.35\textwidth]{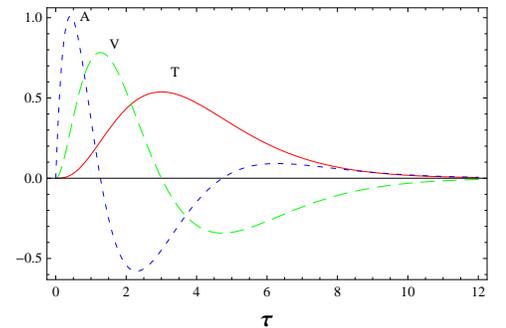}
\end{center}
\caption{\label{tva} The tension transients $T$. The first derivative of tension is $V=\partial_{\tau}T$. The second derivative of tension is $A=\partial^2_{\tau}T$. All the three curves are quit similar to those in Fig. 1 E of Ref. \cite{Gabel}. } \vspace{-0.2cm}
\end{figure}

We take the same quantum Hamiltonian Eq. (\ref{Hkk}) of one dimensional chain of quantum two-level particles for modeling skeletal muscle,
\begin{eqnarray}
H=H_{0}+\sum_{k}(g_{ab}|a_{k}\rangle\langle{b}_{k}|+g_{ba}|b_{k}\rangle\langle{a}_{k}|).
\end{eqnarray}
The exactly solvable Hamiltonian $H_{0}$ is given by Eq. (\ref{H0}). A quantized electric field motivated the hopping between the two levels,
\begin{equation}
g_{ab}(t)=p_{_{0}}E(t)=p_{_{0}}[d(t)+d^{+}(t)],
\end{equation}
where $d$ is annihilation operator of photons. Applying the conventional rotating wave approximation, the total Hamiltonian reads,
\begin{equation}\label{Hdd}
H=H_{0}+p_{_{0}}\rho_{ab}d+p_{_{0}}\rho^{\ast}_{ab}d^{+},
\end{equation}
where 
\begin{equation}
\rho_{ab}=\sum_{k}|a_{k}\rangle\langle{b}_{k}|,\;\;\rho_{ba}=\sum_{k}|b_{k}\rangle\langle{a}_{k}|.
\end{equation}
Hamiltonian Eq. (\ref{Hdd}) shares the very similar formulation with the general Hamiltonian for a coherent state\cite{Mehta}. 
If there is no special projection to certain number of photon, we would always have a coherent state for photons. However, one must project the coherent 
state to a fixed number of photons to get similar curve to the tension-time of skeletal muscle. 
We denote the off-diagonal density operator $\rho_{ab}$ by a brief symbol $\rho=\frac{i}{\hbar}\rho_{ab}$. The Schr$\ddot{o}$dinger equation, 
\begin{equation}
|\psi(t+\Delta{t})\rangle=|\psi(t)\rangle-i\hbar^{-1}H\Delta{t}|\psi(t)\rangle,
\end{equation}
may be expressed by Fock states,
\begin{equation}\label{n(t)}
|n,t+\Delta{t}\rangle=|n,t\rangle-i\frac{{\Delta}t}{\hbar}[\rho\sqrt{n}|n-1,t\rangle-\rho^{\ast}\sqrt{n+1}|n,t\rangle].
\end{equation}
Here we have input the annihilation and creation rule of photon operator, 
\begin{equation}
d^{+}\mid{n}\rangle=\sqrt{n+1}\mid{n+1}\rangle,\;\;\;d\mid{n}\rangle=\sqrt{n}\mid{n-1}\rangle.
\end{equation}
Eq. (\ref{n(t)}) is equivalent to a radioactive decay equation from which one can get the standard Poissonian distribution\cite{Lax}. Physicist define coherent state as the eigenstate of annihilation operator $d$. The coherent state with respect to Hamiltonian 
Eq. (\ref{Hdd}) is $|\rho{t}\rangle$,
\begin{equation}
|\rho{t}\rangle=\exp[\rho^{\ast}{t}d^{+}-{\rho}{t}d]|0\rangle, \;\;\;\;\;\;   d|\rho{t}\rangle=\rho{t}|\rho{t}\rangle. 
\end{equation}
The general expression of wave function as solution of  Schr$\ddot{o}$dinger equation is $|\psi(t)\rangle=\exp[{\frac{i}{\hbar}H_{0}t}]|\rho{t}\rangle.$ 
Projecting the coherent state $|\rho{t}\rangle$ to Fock space gives us Possonian distribution, 
\begin{equation}
|\rho{t}\rangle=\exp[{-\frac{1}{2}\rho^{\ast}\rho{t}^2}]\sum_{n=0}^{\infty}\frac{(\rho{t})^{n}}{\sqrt{n!}}|n\rangle.
\end{equation}
This coherent state is the superposition of many photon states. A photon is the electromagnetic radiation of quantum particles. 
Usually the electromagnetic radiation of human muscle is in the infrared region at a wavelength of tens of microns. 
The detector of electric signal in the muscle is metallic needle. We assume the detector of photon has an optimal window which 
only allow finite number of photon pass, it can not capture all the photons. Then the projection of coherent 
state to this optimal photon number state is what the experiment measured. This electric signal is proportional to the amplitude of the optimal photon state,
\begin{equation}
{T(n,t)}=\arrowvert\langle{n}|\rho{t}\rangle\arrowvert
=\frac{(\rho{t})^{(n-n_0)}}{\sqrt{(n-n_0)!}}\exp[{-\arrowvert\rho{t}\arrowvert^2/2}].
\end{equation}
The experiment of Ref. \cite{Gabel} only reported one complete tension-time course(Fig. 1 E in Ref. \cite{Gabel}). The tension transient of Fig. 1 E in Ref. \cite{Gabel} can be theoretically fitted by 
\begin{equation}
{T(6,\tau)}=\frac{(\rho{\alpha\tau})^{3}}{6!}\exp[{-\arrowvert\rho{\alpha\tau}\arrowvert}],  \;\;\;\alpha\tau=t^2,
\end{equation}
where $\alpha$ is an adjustable parameter to rescale the unit time scale. One can choose different $\alpha$ to get different amplitude for the curve. 
Both the second and the third derivative of ${T(6,\alpha\tau)}$ looks very similar to experiment curve(Fig. \ref{tva}), 
except the minor oscillating tail of the curve. This oscillating tail is most likely due to fluctuating experimental environment 
or the poor quality of equipment in 1968. Because the experimental curves in 1977 of Ref. \cite{Steiger} rarely showed similar wavy 
tails.

\section{The self-coupled Hamiltonian for the microscopic representation of velocity}

There exist relative sliding motion between two filaments. The total number of particles described by the quantum Hamiltonian is sandwiched between the two filament. For an almost infinitely large system, the sliding motion of filament acts as environmental parameters. So we summarize the sliding velocity as a constant parameter in the model. In the quantum Hamiltonian of the one dimensional chain model, the total probability of 
being in the two quantum states is conserved. The quantum density operator measures the probability for a particle being in 
certain state. For the two-level model, the density operator follow $\rho_{aa}+\rho_{bb}=1$. For the three-level model, 
the density operators of three levels obey $\rho_{aa}+\rho_{bb}+\rho_{cc}=1$.

When the one dimensional quantum chain has finite length, the sliding velocity of filament may depend on 
the internal states of motor molecule that drags the filament. The total number of particles sandwiched between 
the two filaments is proportional to the length of the overlapping region. When the myosin molecules between the two filaments are excited, 
they becomes active to drag the filament. 
However if the molecules falls out of the overlap region, even if they are excited, there is no active bond connecting 
the two filaments. We call the motor molecules out of the overlap region as unemployed molecules. 
The molecules within the overlap region are named after employed molecules. 
The mutual sliding movement make the quantum chain an open system. 
The total number of employed particles are increasing during contraction.

A faster sliding velocity means the total number of employed particle increase or decrease faster, so we define 
the mutual sliding velocity as the speed of increasing employed molecules plus a initial velocity $v_{c}$, 
\begin{equation}
v(t)=\frac{dN_{e}}{dt}+v_{c}.
\end{equation}
The absolute value of $v(t)$ is equivalent to the decrease of unemployed molecules. We assume the initial velocity $v_{c}$ is constant. The initial velocity maybe induced by chemical or biological effect. For the simplest case, one may take $v_{c}=0$. The time dependent sliding velocity for the quantum chain of two-level particles can be expressed in terms of density operator,
\begin{equation}
v=\partial_{t}\sum_{j}({\rho}_{aa}^j+{\rho}_{bb}^j) +v_{c}, \;\;\;{\rho}_{aa}^j+{\rho}_{bb}^j\neq{const}.
\end{equation} 
According to Heisenberg equation, $\dot{\rho}=-{i}[H,\rho]$, this velocity is given by the commutator equation of Hamiltonian operator,
\begin{equation}
v=-i\sum_{j}[H,{\rho}_{aa}^j]-i\sum_{j}[H,{\rho}_{bb}^j]+v_{c}.
\end{equation}
Usually a quantum Hamiltonian is expressed by a functional of many operators which does not include the Hamiltonian operator itself. Here the expanding sequence of operators for Hamiltonian Eq. (\ref{Hamiltonian1}) includes a velocity which is a functional of Hamiltonian operator itself. 
The explicit formulation of Hamiltonian Eq. (\ref{Hamiltonian1}) including the Heisenberg equation of velocity reads, 
\begin{eqnarray}
H&=&\sum_{i}\omega_{i}^{a}|a\rangle_{i,i}\langle{a}|+\omega_{i}^{b}|b\rangle_{i,i}\langle{b}|+g_{ab}|a\rangle_{i,i}\langle{b}|
+g_{ba}|b\rangle_{i,i}\langle{a}|\nonumber\\
&+&\frac{1}{s}{(1-|a\rangle_{i}\langle{a}|_{i})H}(|a\rangle_{i}\langle{a}|_{i+1}-|a\rangle_{i+1}\langle{a}|_{i})\nonumber\\
&+&\frac{1}{s}{(1-|b\rangle_{i}\langle{b}|_{i})H}(|b\rangle_{i}\langle{b}|_{i+1}-|b\rangle_{i+1}\langle{b}|_{i}),\nonumber\\
&+&\frac{iv_{c}}{s}[ |a\rangle_{i}\langle{a}|_{i+1}+|b\rangle_{i}\langle{b}|_{i+1}
-|a\rangle_{i+1}\langle{a}|_{i}-|b\rangle_{i+1}\langle{b}|_{i}].\nonumber
\end{eqnarray}
We may call this Hamiltonian a self-coupled Hamiltonian. Transforming the self-coupled Hamiltonian into a conventional Hamiltonian must follow the algebra rules of operating matrix. 
We denote the projection of Hamiltonian to local eigen states as,
\begin{eqnarray}
&&\langle{a}|_{i}H|a\rangle_{i}=\omega_{i}^{a},\;\;\;\;\langle{a}|_{i}H|a\rangle_{i+1}=t_{i,i+1}^{a},\nonumber\\
&&\langle{b}|_{i}H|b\rangle_{i}=\omega_{i}^{b},\;\;\;\;\;\langle{b}|_{i}H|b\rangle_{i+1}=t_{i,i+1}^{b}.
\end{eqnarray}
$t_{i,i+1}^{a}$ and $t_{i,i+1}^{b}$ is the hopping rate between two neighboring sites. 
$\omega_{i}^{a}$ and $\omega_{i}^{b}$ is the local eigen energy of the particle at the $i$th site. After some algebra, we obtained the conventional expansion of 
the self-coupled Hamiltonian,
\begin{eqnarray}\label{HamiG}
H&=&G\sum_{i}[\omega_{i}^{a}+{t_{i,i+1}^{a}}s^{-1}]|a\rangle_{i,i}\langle{a}|\nonumber\\
&+&[\omega_{i}^{b}+{t_{i,i+1}^{b}}{s}^{-1}]|b\rangle_{i,i}\langle{b}|\nonumber\\
&+&g_{ab}|a\rangle_{i,i}\langle{b}|+g_{ba}|b\rangle_{i,i}\langle{a}|+[\frac{iv_{c}}{s}-\frac{\omega_{i}^{b}}{s}]|b\rangle_{i}\langle{b}|_{i+1}\nonumber\\
&-&\frac{iv_{c}}{s}[|a\rangle_{i+1}\langle{a}|_{i}+|b\rangle_{i+1}\langle{b}|_{i}]+[\frac{iv_{c}}{s}-\frac{\omega_{i}^{a}}{s}]|a\rangle_{i}\langle{a}|_{i+1},\nonumber
\end{eqnarray}
where $G$ is the renormalization operator,
\begin{eqnarray}
\frac{s}{G}={s-\sum_{i}[|a\rangle_{i}\langle{a}|_{i+1}+|b\rangle_{i}\langle{b}|_{i+1}
-|a\rangle_{i+1}\langle{a}|_{i}-|b\rangle_{i+1}\langle{b}|_{i}]}.\nonumber
\end{eqnarray}
We can map the self-decoupled Hamiltonian expressed by density operator into a more familiar equation for quantum field theory. Each of the two quantum states 
is represented by a boson operator,
\begin{equation}
\langle{\alpha}_{i}|=\alpha_{i},\;\;|\alpha\rangle_{i}=\alpha_{i}^+, \;\;\alpha=a,b.
\end{equation}
Notice here we study an open finite system, the total particle number is not conserved, i.e., $a^{+}_{i}a_{i}+b^{+}_{i}b_{i}\neq{const}$. 
The decoupled Hamiltonian is simplified into a two-component boson system, 
\begin{eqnarray}\label{HamiG1}
H&=&G\sum_{i}[\omega_{i}^{a}+{t_{i,i+1}^{a}}s^{-1}]a^{+}_{i}a_{i}
+[\omega_{i}^{b}+{t_{i,i+1}^{b}}{s}^{-1}]|b^{+}_{i}b_{i}\nonumber\\
&+&[\frac{iv_{c}}{s}-\frac{\omega_{i}^{a}}{s}]a^{+}_{i}a_{i+1}
+[\frac{iv_{c}}{s}-\frac{\omega_{i}^{b}}{s}]b^{+}_{i}b_{i+1},\nonumber\\
&+&g_{ab}a^{+}_{i}b_{i}+g_{ba}b^{+}_{i}a_{i}-\frac{iv_{c}}{s}[a^{+}_{i+1}a_{i}+b^{+}_{i+1}b_{i}],\nonumber\\
G&=&\frac{1}{1-\frac{1}{s}\sum_{i}[a^{+}_{i}a_{i+1}+b^{+}_{i}b_{i+1}
-a^{+}_{i+1}a_{i}-b^{+}_{i+1}b_{i}]}.
\end{eqnarray}
This nonlinear Hamiltonian has no resemblance in conventional quantum field theory due to the renormalization operator $G$. 
If we assume ${1}/{s}\ll1$, then the operator $G$ is approximated by a constant $G\approx1$. 
Such a approximation is not always appropriate, as the hopping step size $s$ is usually very small. For the other extreme case of 
${1}/{s}\gg1$, the constant term $1$ in the nominator of $G$ can be neglected,
\begin{equation}
G\approx\frac{-s}{\sum_{i}a^{+}_{i}a_{i+1}+b^{+}_{i}b_{i+1}
-a^{+}_{i+1}a_{i}-b^{+}_{i+1}b_{i}}.
\end{equation}
The key rules for handling this inverse quantum operator $G$ is the elementary commutator relation between a matrix operator and the inverse of another operator,
\begin{equation}
[\hat{a}_{i},\frac{1}{\hat{b}_{j}}]=\hat{C}_{ij}. 
\end{equation}
We can assume if $i\neq {j}$, $\hat{C}_{ij}=0$, this is a physical assumption since two elementary bosonic operators are sitting at different lattice sites. While for $i={j}$, $\hat{C}_{ij}$ depends on the specific formulation of the 
mattix $\hat{a}_{i}$ and $\hat{b}_{i}$. We know only one universal relation, $\hat{a}_{i}\frac{1}{\hat{a}_{i}}=1$, i.e., $[\hat{a}_{i},\frac{1}{\hat{a}_{i}}]=0$. Physically speaking, $\hat{a}_{i}$ 
annihilate one particle at the $i$th site. Its dual operator $\frac{1}{\hat{a}_{i}}$ is not the conjugate operator $\hat{a}^{\dag}_{i}$. More over, it is unknown how to decompose the commutator between one elementary operator and the inverse of many composite operators, such as, 
\begin{equation}
[\hat{a}_{i},\frac{1}{\sum_{j}\hat{b}_{j}}]=\hat{C}(a,b). 
\end{equation}
Maybe we should first find out how to map the inverse operator into a series expansion of the conventional polynomial operators.

In fact, quantum operator provides a natural description of directed molecule motors. 
We can define a walking operator $w^{\pm}$, when it operates on a spatial coordinate, it walks one step forward or backward, 
\begin{equation}
w^{+}|x\rangle=|x+s_{0}\rangle,\;\;\;\;w^{-}|x\rangle=|x-s_{0}\rangle.
\end{equation}
$s_0$ is the unit step size. A directed molecule motor is characterized by only annihilation operator or creation operator, 
\begin{equation}
(w^{+})^{m}|x\rangle=|x+ms_{0}\rangle,\;\;\;\;(w^{-})^{m}|x\rangle=|x-ms_{0}\rangle.
\end{equation}
The velocity of the motor is denoted as the number of step $m$ within one unit time interval. For the sliding filaments of muscle fibre, $m$ has a upper bound and lower bound. This is an open system in which the Hamiltonian is only consists of either annihilation operator or creation operator. This open physical system is strongly coupled with another chemical system or biological system.

\section{Conclusion}

When single muscle fibre is teased away from the whole muscle, it is no longer a living organism. In the eye of a physicist, one muscle fibre is a long cylindrical material with complex internal structure. The polymers, chiral proteins and ions inside the fibre are properly organized both in space and functionality. An input electric signal induced the deformation of molecule motors, which in turn shortens the fibre. Generally I believe the electromagnetic interaction(or in other words chemical bonding) determines the equilibrium conformation of myosin molecules. The stochastic fluctuation from thermal environment only results in some minor modifications of local structure, even though sometimes this minor modification plays a crucial role. One myosin motor molecule can be simplified as a giant quantum particle with finite quantum states. Each state corresponds to one stable molecule configuration. The input electric field induce the hopping among different states.

We mathematically model a muscle fibre as a one dimensional chain of quantum particles representing myosin motor molecules. 
Besides the two quantum states, exited state and ground state in analogy of attached state and detached state of classical model, 
an external quantum state of photon is introduced to couple the dynamics of one quantum particle with another. 
The mutual sliding between two filaments is included in the directed hopping operator of a quantum Hamiltonian. 
Quantum dynamics of this quantum Hamiltonian system gives us similar force-velocity relation to the classical model for the quick release case. 
The force-velocity relation for the case of slow release and unstable states is beyond the capable description of classical mechanics model. However using this quantum Hamiltonian, we can derived a rather brief force-velocity relation for the case of slow release and unstable states. The quantum equation of motion suggests that the mutual sliding is a source of the decay of excited states. Another source of decay is vacuum fluctuation or thermal fluctuation, we summarize them into a spontaneous decay process. The mutual sliding movement means breaking bonds and establishing new bonds between two filaments. The ratio between the life time of the sliding induced decay and spontaneous decay determines the behavior of force-velocity relation. For a quick release, the sliding decay is much larger than spontaneous decay. For a slow release, the spontaneous decay dominates. Both the experimental measurement of quick release and slow release only records the behavior of steady states solution. For the non-steady state, the force-velocity relation is an oscillating function of time. There is a difference between the frequency of force oscillation and the input frequency of electric stimulus. This theoretical conclusion is consistent with biological observation---some insect flight muscle indeed can oscillate more rapidly than the frequency of input nervous impulse.

We proposed the mathematical definition of the tension of a muscle fibre as the total number of particles in the excited states. Thus tension transients is described as the evolution of the density operator of the excited states. Based on this hypothesis, we apply the conventional quantum two-level model and three-level model to check the tension transients course of muscle fibre. The quantum two-level model reproduced most of tension-time courses of cardiac muscle under stretch activation. The three-level model can theoretically generate most curves of the tension transient of insect flight muscle. The quantum three-level model also produced some tension-time course of rabbit psoas muscle for which the two-level model can not account. The key difference between three-level model and two-level model is that three-level model takes into account of the induced decay due to metastable state while the two-level model does not. The quantum three-level model can also generate the tension transients of a skeletal muscle fibre under stimulus of electric pulse. However we checked the tension transients of skeletal muscle from a different point of view of quantum coherent state. The tension, the first derivative of tension and the second derivative of tension are all coincide with the third order projection of coherent states. It is not fully understood in physics why it projects to the third order, maybe there are better ways to model skeletal muscle.

In the muscle stimulus experiment almost 35 years ago\cite{Steiger}, the tension-time course varies according to the concentration of $Ga^{++}$ and $Mg^{++}$. In the quantum chain model, those mathematical parameters can varies to produce similar curves to the experiment data. If there are too many theoretical parameters, it is big challenge to understand the physical or chemical relation between the concentration of $Ga^{++}$ and $Mg^{++}$ and those parameters. Fortunately there are only three effective parameters in the quantum model related to the muscle tension transient : the two decay rates of excited states and the frequency difference between the two excited states. The decay rates controls the long time behavior of tension transient. The frequency difference introduces wavy oscillations upon the conventional exponential decay or exponential increase. It needs a series of systematic experiments to 
find out how ions control the three parameters, the existed experiment data is not enough. 
One experiment fact is that if solution does not contain $Ga^{++}$ ions, delayed tension changed disappeared, if the $Ga^{++}$ and $Mg^{++}$ are added, distinctive tension transients were observed\cite{Steiger}. It seems as if $Ga^{++}$ ion drive the myosin motor molecule into metastable conformation and transform the molecule into a three-level quantum particle. While $Mg^{++}$ seems like to fix the conformation of myosin molecules so that they are invariant under stretch activation. Maybe the ratio between the concentration of $Ga^{++}$ and $Mg^{++}$ is a possible controller of the two decay rate of the two excited states. Both theoretical and experimental works are required to solve the puzzle.

\section{Acknowledgment} 

I did not find the contact information of the author of Ref \cite{Steiger} to ask for the permission of using the experimental curves in my article. 
If the reader want to see the experimental curve but can not find the old Ref. \cite{Steiger} in 1977, please sent me an E-mail. I will ask the electronic library 
in Max-Planck-Institute for the Physics of Complex Systems for help.

\section{Appendix}

\appendix{}

\section{The derivation of force velocity relation for the case of quick release}\label{quickrelease}

I present the detail calculations of force-velocity equation here.  The assumptions and approximations used in calculation are explained at every step. I first show how I derived the differential equations from the quantum chain model. Then I add the general procedure for solving similar differential equations in Ref. \cite{Sargent} for the benefit of readership. The assumption used in solving the very similar differential equations have more detail explanation in laser physics theory\cite{Sargent}.

For a quantum particle, the probability of being in a quantum state is given by the square of the norm of $|\alpha\rangle$, i.e., $\rho=\mid|\alpha\rangle\mid^2$. This probability is well defined by density operator,  ${\rho}_{\alpha\alpha}=|\alpha\rangle\langle{\alpha}|$. For a two level system, there are four density operators, 
\begin{eqnarray}
{\rho}_{aa}^i&=&|a\rangle_{i,i}\langle{a}|,\;\;\;\;
{\rho}_{bb}^i=|b\rangle_{i,i}\langle{b}|,\nonumber\\
{\rho}_{ab}^i&=&|a\rangle_{i,i}\langle{b}|,\;\;\;\;
{\rho}_{ba}^i=|b\rangle_{i,i}\langle{a}|.\nonumber\\
\end{eqnarray}
The diagonal density operator satisfy the constrain, 
\begin{equation}
1=\frac{1}{2}[|a\rangle_{i,i}\langle{a}|+|b\rangle_{i,i}\langle{b}|], 
\end{equation}
The equation of motion for these density operator is
\begin{equation}\label{dtrho}
\dot{\rho}=-\frac{i}{\hbar}[H,\rho]=-\frac{i}{\hbar}[H\rho-\rho{H}]. 
\end{equation}
The Hamiltonian of the one dimensional chain model is
\begin{eqnarray}
H&=&\sum_{i}\omega_{i}^{a}|a\rangle_{i,i}\langle{a}|+\omega_{i}^{b}|b\rangle_{i,i}\langle{b}|+g_{ab}|a\rangle_{i,i}\langle{b}|
+g_{ba}|b\rangle_{i,i}\langle{a}|\nonumber\\
&+&\frac{iv}{s}[ |a\rangle_{i}\langle{a}|_{i+1}+|b\rangle_{i}\langle{b}|_{i+1}
-|a\rangle_{i+1}\langle{a}|_{i}-|b\rangle_{i+1}\langle{b}|_{i}].\nonumber
\end{eqnarray}
Substituting this Hamiltonian and density operator into Eq. (\ref{dtrho}), one may get the equation of motion of the probability density in different states. Here we take ${\rho}_{aa}^i$ as an example, 
\begin{eqnarray}\label{raa}
{i\hbar}\dot{\rho}_{aa}^i&=&-\frac{iv}{s}\{[|a\rangle_{i+1}-|a\rangle_{i}]\langle{a}|_{i}+|a\rangle_{i}[\langle{a}|_{i+1}-\langle{a}|_{i}]\}\nonumber\\
&+&g_{ba}|b\rangle_{i,i}\langle{a}|-g_{ab}|a\rangle_{i,i}\langle{b}|-2\frac{iv}{s}|a\rangle_{i}\langle{a}|_{i}.
\end{eqnarray}
If the lattice spacing $s$ is very small, we summarize the first term of Eq. (\ref{raa}) into a continuum limit, 
\begin{equation}
\frac{1}{s}\{[|a\rangle_{i+1}-|a\rangle_{i}]\langle{a}|_{i}+|a\rangle_{i}[\langle{a}|_{i+1}-\langle{a}|_{i}]\}
=\partial_{x}{\rho}_{aa}
\end{equation}
Here $\hbar$ is set to $\hbar=1$ for convenience. Then Eq. (\ref{raa}) becomes more compact,
\begin{eqnarray}
(\partial_{t}+{v}\partial_{x}){\rho}_{aa}^i&=&n^0_{a}+{i}g_{ab}{\rho}_{ab}^i{-i}g_{ba}{\rho}_{ba}^i-2\frac{v}{s}{\rho}_{aa}^i.
\end{eqnarray}
Following a similar calculation, we get the equation of ${\rho}_{bb}^i$ and ${\rho}_{ab}^i$,
\begin{eqnarray}
(\partial_{t}+{v}\partial_{x}){\rho}_{bb}^i&=&n^0_{b}+{i}g_{ba}{\rho}_{ba}^i{-i}g_{ab}{\rho}_{ab}^i-2\frac{v}{s}{\rho}_{bb}^i,
\end{eqnarray}
\begin{equation}\label{rab}
(\partial_{t}+{v}\partial_{x}){\rho}_{ab}^i=
{i}g_{ab}({\rho}_{bb}^i-{\rho}_{aa}^i)
{-i}\Delta\omega_{i}{\rho}_{ab}^idraft
-2\frac{v}{s}{\rho}_{ab}^i,
\end{equation}
where $\Delta\omega_{i}=(\omega_{i}^{a}-\omega_{i}^{b})$.
Since ${\rho}_{ab}^{\ast}={\rho}_{ba}$. The equation of motion for ${\rho}_{ba}$ is just the conjugate of ${\rho}_{ab}$. 
Mathematical theory told us, if the equation bear the form of 
\begin{equation}
(\frac{\partial}{\partial{t}}+v\frac{\partial}{\partial{x}})f(x,t,v)=g(x,t,v),
\end{equation}
the solution is formulated as 
\begin{equation}
f(x,t,v)=\int_{-\infty}^{t}dt'g(x',t',v),\;\;\;x'=x-v(t-t').
\end{equation}
The above equations of motion for density operator has exactly this kind of form. The solutions can be obtained by analogy of the calculation in quantum optics\cite{Sargent}. Within the rotating wave approximation, the coupling coefficient are   
\begin{equation}
g_{ab}(t')=\frac{p_{_{0}}}{2}\sum_{n}E_{n}(t')e^{-i(f_{n}t'+\phi_{n})}U_{n}(z'),
\end{equation}
Integration of Eq. (\ref{rab}) gives out 
\begin{eqnarray}
&&{\rho}_{ab}^i(z,v,t)={i}\int_{-\infty}^{t}dt'\exp[-(i\omega+2\frac{v}{s})(t-t')]\nonumber\\
&&{\times}g_{ab}(z',t')({\rho}_{bb}^i(z',v,t')-{\rho}_{aa}^i(z',v,t')).
\end{eqnarray}
Following the assumptions of laser theory\cite{Sargent}, the population difference ${\rho}_{bb}^i(z',v,t')-{\rho}_{aa}^i(z',v,t')$, the electric field mode amplitude $E_{n}$ and the phase $\phi_{n}$ vary little in the time ${s}/({2v})$, then ${\rho}_{ab}$ is rewritten as, 
\begin{eqnarray}
&&{\rho}_{ab}^i=-\frac{1}{2}{i}\frac{p_{_{0}}}{\hbar}E_{n}\exp[-i(f_{n}t'+\phi_{n})]\nonumber\\
&&\left\lbrace \int_{-\infty}^{t}dt'U_{n}(z')\exp[-[i(\omega-f_{n})+\frac{2v}{s}](t-t')]\right\rbrace \nonumber\\
&&{\times}[{\rho}_{bb}^i(z',v,t')-{\rho}_{aa}^i(z',v,t')].
\end{eqnarray}
$U_{n}(z')$ is the normal-mode function in space. For the one dimensional chain model, $U_{n}(z')$ is the distribution of electric field along the length direction of the chain. As the electric field is oscillating in space, one usually take sinusoidal $z$ dependence: $U_{n}(z')=\sin[k_n{z'}]$. As all know, $\exp[ik_n{z}]=\cos[k_n{z}]{\pm}i\sin[k_n{z}]$, $U_{n}(z')$ is decomposed into the sum of two opposite waves. For a symmetric velocity distribution $W(-v)=W(v)$,  the 
quick $\cos[k_n{z}]$ terms are canceled in the integration, it finally yields $U^2_{n}(z)={1}/{2}$. The density operator ${\rho}_{ab}$ now reads,  
\begin{equation}\label{rabDD}
{\rho}_{ab}^i=-\frac{i}{4}\frac{p_{_{0}}}{\hbar}E_{n}e^{-i(f_{n}t+\phi_{n})}
({\rho}_{bb}^i-{\rho}_{aa}^i)(D_{+}+D_{-}),
\end{equation}
where $D_{\pm}$ is the propagator function, 
\begin{equation}
D_{\pm}=\frac{1}{2{v}/{s}+i(\omega-f_{n}{\pm}kv)},
\end{equation}
where $k=\omega/c$, $c$ is the speed of light. Substituting Eq. (\ref{rabDD}) into the equation of motion of ${\rho}_{aa}$ and ${\rho}_{bb}$, one obtain the rate equation, 
\begin{eqnarray}
\partial_{t}{\rho}_{aa}^i&=&n^0_{a}-R(v)({\rho}_{aa}^i-{\rho}_{bb}^i)-2\frac{v}{s}{\rho}_{aa}^i,\nonumber\\
\partial_{t}{\rho}_{bb}^i&=&n^0_{b}+R(v)({\rho}_{aa}^i-{\rho}_{bb}^i)-2\frac{v}{s}{\rho}_{bb}^i.
\end{eqnarray}
where 
\begin{eqnarray}\label{R(v)}
R(v)=-\frac{1}{8}
\frac{p^2_{_{0}}E^2_{n}}{\hbar^2}\frac{s}{2v}[G_{-}+G_{+}],
\end{eqnarray}
$G_{\pm}$ has the similar form of Lorentzian function  
\begin{equation}
G_{\pm}=\frac{4{v}^2/{s}^2}{4{v}^2/{s}^2+(\omega-f_{n}{\pm}kv)^2}.
\end{equation}
For the steady state,  $\partial_{t}{\rho}_{aa}^i=0$, $\partial_{t}{\rho}_{bb}^i=0$, the pair of rate equation leads to
\begin{equation}\label{fv0}
{\rho}_{aa}^i-{\rho}_{bb}^i=\frac{{\Delta}n(z)}{1+R\frac{s}{v}}.
\end{equation}
Here ${\Delta}n(z,v,t)$ is the initial population difference in the absence of field oscillation, 
\begin{equation}
{\Delta}n(z,v,t)=\frac{s}{2v}[n^0_{a}(z,v,t)-n^0_{b}(z,v,t)].
\end{equation}
Eq. (\ref{fv0}) gives the relation between population difference and decay coefficient. In our one dimensional chain model, the decay coefficient is just the linear function of velocity. So Eq. (\ref{fv0}) is actually the force-velocity relation,  
\begin{equation}
({\rho}_{aa}^i-{\rho}_{bb}^i)({v+R{s}})=\frac{s[n^0_{a}(z,v,t)-n^0_{b}(z,v,t)]}{2}. 
\end{equation}
The rate function $R(v)$ is a complex function of velocity. At the resonance points $\omega-f_{n}={\pm}kv$,  $R(v)$ gained maximal value.

\end{document}